\newcommand\papertitle{How bad could it be? Modelling the 3D complexity of the polarised dust signal using moment expansion}

\documentclass{aa}

\makeatletter
\renewcommand*\aa@pageof{, page \thepage{} of \pageref*{LastPage}}
\makeatother
\usepackage{bbm}
\usepackage{graphicx}
\usepackage{txfonts}
\usepackage{url}
\usepackage{natbib}
\bibpunct{(}{)}{;}{a}{}{,} 
\usepackage{color}
\usepackage{colortbl} 
\usepackage[dvipsnames]{xcolor}
\usepackage{enumitem}
\usepackage{tipa}
\usepackage{booktabs}
\usepackage[switch]{lineno}

\usepackage[breaklinks, colorlinks, citecolor=blue]{hyperref}
\usepackage{subfigure}
\usepackage{epsfig}
\usepackage{multirow}
\usepackage{array}

\usepackage{url}
\usepackage{psfrag}
\usepackage[T1]{fontenc}
\usepackage{rotating}
\usepackage{soul}
\usepackage{wrapfig}
\usepackage{tikz}

\makeatletter
\def\env@cases{
  \let\@ifnextchar\new@ifnextchar
  \left\lbrace
  \def\arraystretch{1.2}
  \array{l@{}l@{}}
}
\makeatother

\definecolor{mygreen}{RGB}{104,198,107}
\definecolor{myred}{RGB}{252,137,125}
\definecolor{myyellow}{RGB}{252,225,126}
\definecolor{mygrey}{RGB}{215,215,215}

\newcommand{\review}[1]{{#1}}
\newcommand{\reviewcorrect}[2]{{#2}}

\definecolor{darkred}{rgb}{0.55, 0.0, 0.0}

\def\barP{\overline{\mathcal{P}}_\nu\left(\overline{\betad},\overline{\tempd}\right)}
\def\Walphap{\mathcal{W}_\alpha^p}

\def\i{\mathbbm{i}}
\def\ipix{{i_{\rm pix}}}

\def\d{\rm d}

\def\planck{\textit{Planck}}

\def\dten{{\tt d10}}

\def\dtwelve{{\tt d12}}

\def\betad{\beta}

\def\tempd{\mathcal{T}}

\def\polnu{\varepsilon_\nu}

\begin{document}

\title{\papertitle}

\offprints{\url{lvacher@sissa.it}}
\authorrunning{Vacher et al.}
\titlerunning{}

\author{L. Vacher\inst{\ref{SISSA},\ref{INFN},\ref{IFPU}}
\and 
A. Carones\inst{\ref{SISSA},\ref{INFN},\ref{IFPU}} 
\and J. Aumont\inst{\ref{IRAP}}
\and J. Chluba\inst{\ref{Manch}}
\and 
N. Krachmalnicoff\inst{\ref{SISSA},\ref{INFN},\ref{IFPU}}
\and
C. Ranucci\inst{\ref{SISSA},\ref{INFN},\ref{IFPU}}
\and\\
M. Remazeilles\inst{\ref{IFC}}
\and
A. Rizzieri\inst{\ref{Oxford}}
}
\institute{
International School for Advanced Studies (SISSA), Via Bonomea 265, 34136, Trieste, Italy \label{SISSA}
\and
 Istituto Nazionale di Fisica Nucleare (INFN), Sezione di Trieste, via Valerio 2, 34127 Trieste, Italy\label{INFN}
\and
\review{Institute for Fundamental Physics of the Universe (IFPU), Via Beirut, 2, 34151 Grignano, Trieste, Italy} \label{IFPU}
\and
Institut de Recherche en Astrophysique et Planétologie (IRAP), Universit\'e de Toulouse, CNRS, CNES, UPS, 9 Av. du Colonel Roche, 31400 Toulouse, France\label{IRAP}
\and 
Jodrell Bank Centre for Astrophysics, Alan Turing Building, University of Manchester, Oxford Rd M13 9PL, Manchester, United Kingdom \label{Manch}
\and 
Instituto de Fisica de Cantabria (CSIC-UC),
Avda. los Castros s/n, 39005 Santander, Spain
\label{IFC}
\and
Department of Physics, University of Oxford, Denys Wilkinson Building, Keble Road, Oxford OX1 3RH, United Kingdom \label{Oxford}
}

\abstract{The variation of the physical conditions across the three dimensions of our Galaxy is a major source of complexity for the modelling of the foreground signal facing the cosmic microwave background (CMB). In the present work, we demonstrate that the spin-moment expansion formalism provides a powerful framework to model and understand this complexity, and we put special focus on the effects that arise from variations of the physical conditions along each line of sight on the sky. We performed the first application of the moment expansion to reproduce a thermal dust model largely used by the CMB community, demonstrating its power as a minimal tool to compress, understand, and model the information contained within any foreground model. Furthermore, we used this framework to produce new models of thermal dust emission containing the maximal amount of complexity allowed by the current data while remaining compatible with the observed angular power spectra by the \planck{} mission. By assessing the impact of these models on the performance of component separation methodologies, we conclude that the additional complexity contained within the third dimension could represent a significant challenge for future CMB experiments and that different component separation approaches are sensitive to different properties of the moments.}


\maketitle

\section{Introduction \label{sec:introduction}}
Our modern understanding of cosmology relies heavily on the precise characterisation of the cosmic microwave background (CMB) signal. The temperature anisotropies of the CMB have been accurately mapped over the whole sky, validating the canonical $\Lambda$-CDM model as the best-fit model to describe the content and evolution of our Universe \citep{Planck18_I,PlanckCosmo}. Future experiments are now focusing on the fainter polarisation signal of the CMB in order to unveil the earliest phases of our cosmic history. In particular, the standard cosmological model may require a phase of accelerated expansion -- cosmic inflation -- occurring in the very first fraction of a second after the primordial singularity to explain multiple observational facts about the early Universe \citep{inflationhist1,inflationhist2,inflationhist3}. If this event occurred, it generated a gravitational wave background bathing the primordial plasma and led to a characteristic large-scale polarisation pattern -- the primordial $B$ modes -- being imprinted in the CMB   \citep{InflationmodesB3,InflationsmodesB1,InflationsmodesB2}. In the current leading hypothesis, the energy density at the origin of cosmic inflation has to be generated by one or multiple novel fundamental fields beyond those of the standard model of particle physics. Detection of the primordial $B$ modes thus represents one of the most ambitious targets of modern cosmology for the next decades, but it will allow for testing of fundamental physics at energy scales beyond the reach of particle accelerators. The amplitude of the primordial $B$ modes is quantified by the tensor-to-scalar ratio $r$, which is directly related to the energy scale at which inflation occurred. It is constrained by current data to satisfy $r\lesssim 0.03$ (95\% C.L.) \citep{tristram,Galloni2023}. However, contemporary and future missions will probe the value of this parameter at the $10^{-3}$ level in the next decades \citep{SimonsObservatory,Ptep,CMBS4}.

In order to reach such a goal, multiple challenges have to be faced. 
In particular, our own Galaxy produces complex polarised signals -- known as foregrounds -- that contaminate the CMB and are several orders of magnitude brighter than it. Distinguishing the Galactic signal from the cosmological signal requires a delicate procedure known as component separation \citep{Dickinson2009,PlanckCompoSep}. 
As CMB experiments become increasingly sensitive to the primordial signal, they accordingly become more sensitive to the complexity of the foregrounds \citep{Remazeilles_etal_2016}. 
Thus, they require refined component separation methods.

In the frequency range relevant for CMB observations, the polarised Galactic foreground signal is produced by two main sources: synchrotron emission and thermal dust signal. Synchrotron, dominating at low frequencies, is generated by light charged particles (cosmic rays) accelerating and spiralling around the lines of the Galactic magnetic field (GMF). Thermal dust, dominating at high frequencies, is the consequence of stellar light being absorbed and re-emitted by dust grains populating the interstellar medium (ISM) and rotating on themselves perpendicularly to the GMF.
Local physical conditions in the ISM such as temperature, composition, density, and orientation of the GMF vary across the three dimensions of the Galaxy. This fact is supported by both observational and theoretical considerations ranging from the scale of the filaments to the scale of the whole Galaxy \citep{Ferriere, dustacrossMW, Ysard2013, Jaffe2013, vardustdisk2, Planck18_XI,Zelko2022,QUIJOTE2023,Pelgrims2024,Liu2024}.
The variation of these physical conditions is directly traced by the light charged particles and the dust grains present in the ISM and is associated with a 3D variation of their emission properties. The spectral parameters and local polarisation orientation associated to the synchrotron and the thermal dust signals thus change over the sky. As such, any astrophysical observation presents itself as a mixture of different local signals associated with different emission conditions. Here, we refer to this phenomenon as `mixing' \citep{Chluba2017}. Mixing occurs along a single line of sight in the 3D Galaxy, between lines of sight in the plane of the sky (e.g. in the instrumental beam or map pixel), and 
over large patches of the sky when a spherical harmonic transformation is performed. 

Due to the non-linearity and geometric nature of the local polarised foreground emission, the mixing has significant and interdependent consequences on the total observed signal. 
First, as the sum of multiple non-linear functions gives a different function, the total (polarised) intensity will not behave as the local spectral energy distribution (SED) anymore. This phenomenon is known as `SED distortions\ \citep{Chluba2017,Remazeilles2018,Mangilli}.
Then, as the two Stokes parameters $Q$ and $U$ might be distorted differently, the total polarisation angle will rotate with frequency \citep{tassis,Vacher2022b}. Furthermore, mixing also has consequences for the decomposition of the foreground signal into $E$ and $B$ modes. To the spectral rotation of the polarisation angle corresponds a rotation of the $E$ modes into the $B$ modes, leading to a variation of the $E/B$ ratio of the total signal with frequency \citep{Vacher2023}. Finally, because of the dependence of the emission properties on the direction of an observation, it is not possible to infer a map at a given frequency based on one at another frequency using a constant scaling. Consequently, the cross-frequency angular power spectra loose some power. This last phenomenon is referred to as `frequency decorrelation' \citep{Planck2016_XXX,pelgrims2021}. 

Modelling and measuring all of these effects is a priority for both CMB component separation and Galactic science. 
Indeed, mismodelling of these properties could have dramatic consequences on the recovery of the primordial $B$ modes through component separation in the quest for primordial inflation \citep{PlanckandBICEP} or in the quest for a primordial $EB$ correlation
, signature of parity violating cosmic birefringence in the early Universe\footnote{For a discussion on the $EB$ correlation produced by foregrounds and their impact on CMB analysis, see also \cite{Clark2021,Cukierman2022,Diego-Palazuelos2022a,Jost2023,Hervias-Caimapo2024}.} \citep{Diego-Palazuelos2022b}. On the other hand, detecting and understanding the consequences of mixing allows one to probe the variation of the properties of the ISM in the 3D Galaxy \citep{pelgrims2021,Ritacco22,McBride2022}.%

Moment expansion \citep{Chluba2017} models the SED distortions arising from mixing in a minimal and well motivated way through a Taylor expansion of the SED with respect to the spectral parameters, which is a framework that was also applied to the modelling of Sunyaev-Zeldovich signals to separate and compress spectral and spatial information \citep{Chluba2012SZpack}. This approach can be generalisedd to the polarised signal in order to also model the spectral variation of the polarisation angle \citep{Vacher2022b}.
The formalism was further extended at the angular power spectra level in intensity and polarisation \citep{Mangilli,Vacher2023} such that the consequences of mixing can be captured even in harmonic space.
The moments thus represent a very powerful path towards performing component separation of the CMB foregrounds using minimum variance approaches \citep{RemazeillesmomentsILC,Rotti2021, Adak2021,Carones2024} or parametric fitting in pixel or harmonic space \citep{Ichiki2019,Azzoni2020,Vacher2022a,Sponseller2022,Minami2023,Azzoni2023,Wolz2024}.

In this work, we use moments for the first time to understand and build thermal dust simulations containing additional complexity arising in the third dimension from the line of sight variation of the physical conditions. While this exercise can easily be done both in intensity and in polarisation, we focus on polarisation, aiming our discussion on the impact of such complexity on the measurement of the primordial $B$ modes. The new models will be generated as variations of the ones already provided by the \texttt{PySM} framework, which provides full sky simulations of the foreground signals for the CMB community \citep{Thorne,Zonca2021}. Our objective is twofold. First, we seek to demonstrate that the moment expansion can be used to decompose and interpret the \texttt{PySM} models already existing and used by the CMB community. Secondly, we aim to show that the moments provide a simple and powerful tool to inject any desired additional complexity into these models. This last point is of particular interest to the study of the limits of various component separation methods.

In Sect.~\ref{sec:mom-rev}, we give a general presentation of the moment expansion formalism and its ability to model and tackle foreground complexity. In Sect.~\ref{sec:d12}, we quantitatively assess such an ability by reproducing the line of sight complexity of a dust model used by the CMB community through moment expansion. In Sect.~\ref{sec:model-building}, we re-use the moments computed in the previous section to create maximally complex dust models in the limit set by \planck{} data. We then apply basic component separation techniques to these models in order to quantify the impact of the added complexity. We further attempt to interpret this impact in regard to the moment maps contained in the models. Finally, we discuss the limits of our analysis and present our conclusions in Sect.~\ref{sec:conclusion}.

\section{Moments and the third dimension \label{sec:mom-rev}}

\subsection{Mixing and the moment expansion formalism}

We now introduce the moment expansion framework in full generality, as a powerful tool to model complexity coming from the spatial variation of the foreground properties in polarisation.
Locally, the frequency dependent linearly polarised signal emitted by an idealised point $x$ of our Galaxy\footnote{Typically one can think of $x$ as a triplet of coordinates $x=(x_1,x_2,x_3)$ locating an idealised infinitesimal point of emission in the Galaxy (sometimes given by a position vector $\vec{r}$). However, we intend to be more general here and $x$ could label a sub-region of the instrumental beam or the pixel of a sky map.} is characterised by a {canonical spectral energy distribution} (SED), which can be modelled by a frequency dependent complex number $\mathcal{P}_\nu=Q_\nu+\i U_\nu$, where $Q_\nu$ and $U_\nu$ are the Stokes parameters of linear polarisation.
$\mathcal{P}_\nu$ can be written as
\begin{equation}
\mathcal{P}_\nu=\mathcal{A}\polnu(\vec{p}),   
\end{equation}
where $\polnu$ is a real function of the frequency called the emissivity, which characterises the SED of the signal. $\polnu$ depends on a list of $N$ real parameters $\vec{p}=(p_1,p_2,\dots,p_N)$ called the {spectral parameters}. The modulus of $\mathcal{P}_\nu$ is called the {polarised intensity}. $\mathcal{A}$ is a frequency independent complex number playing the role of a {complex amplitude} or {weight}. Amplitudes are usually defined as the value of $\mathcal{P}_\nu$ evaluated at an arbitrary {reference frequency} $\nu_0$ ($\mathcal{A}=\mathcal{P}_{\rm \nu=\nu_0}$). The phase of $\mathcal{A}$ divided by two, $\psi={\rm Arg}(\mathcal{A})/2$ is a frequency independent angle called the {polarisation angle}\footnote{The factor of $1/2$ here translates the spin-2 nature of $\mathcal{P}_\nu$. In simple words, this means that the polarisation signal is identical to itself if rotated by $180^\circ$, as would a headless vector.}.

In the Galaxy, the physical conditions change when considering different points of emission associated to different spatial regions. As such, different points $x$ will be associated with different values of the complex amplitude and the spectral parameters. When an astrophysical observation is performed, the observed signal $\langle\mathcal{P}_\nu\rangle$ will be the sum of multiple local signals contained in a considered spatial region $\Omega$ as
\begin{equation}
\langle\mathcal{P}_\nu\rangle = \int_\Omega \mathcal{A}(x)\polnu(\vec{p}(x))\d x.
\label{eq:mixing}
\end{equation}
Such a phenomenon is referred to as `(spectral) mixing' or `polarised mixing' for the special case of polarisation. Since $\polnu(\vec{p})$ is not linear in $\vec{p}$, the modulus of the total signal $\langle\mathcal{P}_\nu\rangle$  will not scale across frequencies according to $\polnu$ anymore. We talk about `SED distortions'. This integral will also mix different complex weights with functions of frequencies such that the polarisation angle of the observed signal $\langle\psi\rangle= {\rm Arg}(\langle\mathcal{P}_\nu\rangle)/2$ will become a function of frequency. We refer to this effect as `spectral rotation of the polarisation angle'.

Considering any microwave sky observation, mixing can occur in three different ways: i) along the line of sight in the depth of the sky, ii) in between lines of sights inside the instrumental beam and/or the sky pixel or iii) in between lines of sights over large patches of the sky when performing a spherical harmonic transformation. The consequences of ii) and iii) will depend on the instrumental setup as well as the framework used to analyse the data,\footnote{In particular, pixel based component separation methods will not suffer from iii) but will be majorly impacted by ii). This impact will not only be caused by the convolution of the instrumental beam but also by some required under-pixelisation of the data (downgrading) as well as by how the beam is treated (on this last point see \cite{Rizzieri2024}).} while i) is some physical or intrinsic mixing that is always present at some level in the analysis. It is common to distinguish the consequences of i) and ii) by naming them respectively `line-of-sight' and `plane of sky' complexities. However, in practice, it is impossible to create an absolute distinction between these two effects in general and they can all be modelled using the same formalism as both i) and ii) will appear identically in Eq.~\eqref{eq:mixing}. 
In this work, we attempt to avoid this confusion by remaining in the context of building a model on a pixelised sphere and by distinguishing 'intra-pixel' and 'inter-pixel' complexities to describe respectively i) and ii) at a fixed pixel resolution. 

The idea of the moment expansion formalism is to model the complexity arising from mixing by making a common Taylor expansion of the SED with respect to its spectral parameters, for every emission point one averages over in Eq.~\eqref{eq:mixing} \citep{Chluba2017}. The total resulting SED is itself be modelled by a global Taylor-like expansion, called the `moment expansion' or `spin-moment expansion' for its generalisation to polarisation \cite{Vacher2022b}. It takes the form\footnote{\reviewcorrect{Neglecting all additional possible complications as bandpass integration.}{In the presence of bandpass integration and beam chromaticity, this expansion takes a more complex form (for a discussion, see \cite{Chluba2017,Vacher2022b}). We neglect such effects here and assumed ideal bandpasses that behave as Dirac distributions as well as  frequency independent beam shapes.}}
\begin{align}
    \langle \mathcal{P}_\nu \rangle = \overline{\mathcal{P}}_\nu(\bar{\vec{p}})\times\Bigg(1 &+ \sum_i\mathcal{W}_1^{p_i}\partial_{p_i} \polnu|_{\vec{p}=\bar{\vec{p}}} \nonumber\\
    &+ \frac{1}{2}\sum_{i,j}\mathcal{W}_2^{p_ip_j}\partial_{p_i}\partial_{p_j} \polnu|_{\vec{p}=\bar{\vec{p}}} + ...\Bigg),
\label{eq:mom-exp}
\end{align}
where the overall factor $\overline{\mathcal{P}}_\nu=\overline{\mathcal{A}}\polnu(\bar{\vec{p}})$ is the canonical SED with amplitude $\overline{\mathcal{A}}= \int_\Omega \mathcal{A}(x)\d x$\footnote{For reference $\mathcal{A}$ was noted $\mathcal{W}_0$ in \cite{Vacher2022b,Vacher2023} in order to stress its interpretation as a 'zero-th order moment'.} and {pivot spectral parameters} $\bar{\vec{p}}=(\bar{p}_1,\bar{p}_2,\dots,\bar{p}_N)$ around which the expansion is done. $\mathcal{W}_\alpha^{\vec{p}}$ are the so-called $\vec{p}$ spin-moment coefficients of order $\alpha$. This name is justified as it turns out that these coefficients can be understood as the statistical moments of the spectral parameters distribution weighted by the complex amplitudes\footnote{Assuming that each point of emission $x$ is fully characterised by one value of the weight $A(x)$ and a single value for each spectral parameters $\vec{p}(x)$.} as
\begin{equation}
    \mathcal{W}_\alpha^{p_ip_j \cdots p_\alpha} = \frac{\int_\Omega \mathcal{A}(x)(p_i(x)-\bar{p}_i) (p_j(x)-\bar{p}_j)\cdots\cdot(p_\alpha(x)-\bar{p}_\alpha)  \d x}{\int_\Omega \mathcal{A}(x) \d x}.
\label{eq:moments}
\end{equation}
One can interpret the real part of the first order moment as a correction to the pivot around which the expansion is performed.\footnote{It is deeply insightful to consider the correction of the pivot by the complex number $\mathcal{W}_1^{p_i}$ as discussed in \cite{Vacher2022b,Vacher2023}, but here we do not explore such an idea further.} It is hence possible to correct the pivot values as follows:
\begin{equation}
    \bar{p}_i \to \bar{p}'_i =\bar{p}_i + {\rm Re}\left(\mathcal{W}_1^{p_i}\right) = {\rm Re}\left(\frac{\int_\Omega \mathcal{A}(x)p_i(x)\d x}{\int_\Omega \mathcal{A}(x)\d x}\right).
\label{eq:p_bar}
\end{equation}
Modelling the signal as a moment expansion using $\bar{p}'_i$ as a pivot leads to an expansion with a null real part for the first order (${\rm Re}\left(\mathcal{W}_1^{p_i}\right)=0$) and expected to converge within a minimal amount of terms. The term $\mathcal{W}_1^{p_i}$ thus becomes a purely imaginary number that is expected to be the largest contribution to the spectral rotation of the polarisation angle, as further discussed in Appendix~\ref{app:mom-maps}.

The addition of each of the terms in Eq.~\eqref{eq:mom-exp} is expected to model finer and finer distortions of the polarised SED\footnote{In polarisation, the situation is a bit more subtle as the hierarchy between the moments can be broken (i.e. one can have $|\mathcal{W}_{\alpha}^p| < |\mathcal{W}_{\alpha+1}^p|$).}. The term containing $\mathcal{W}_{\alpha}^p$ is referred to as the term of order $\alpha$ and we say that the expansion is performed up to order $\alpha$ when it includes all the terms in Eq.~\eqref{eq:mom-exp} including the order $\alpha$, while all the terms of order greater than $\alpha$ are assumed to be strictly zero.

The formalism to model intensity $\mathcal{I}_\nu$ is absolutely identical as the one presented above, with the weights being real numbers $A$ instead of complex ones, such that locally $\mathcal{I}_\nu = A I_\nu(p)$. Following \cite{Chluba2017}, real intensity moments are noted $\omega^p_\alpha$ to distinguish them from the spin moments $\mathcal{W}^p_\alpha$.

The moment expansion allowed us to model through a set of a few coefficients the complex effects of polarised mixing induced by possibly infinite emission points associated with different spectral properties. In order to account for some complexity contained along the line of sight in the third dimension, one would have to use a set of moment coefficients in each line of sight, or equivalently in every pixel of a map. As such, the information about the complexity along the third dimension within each pixel can be fully captured by a set of maps $(\omega^p_i,\mathcal{W}^p_i)$ and their associated SEDs, informing about the complexity contained both in intensity and polarisation within each pixel.

\begin{figure*}[h!]
    \centering \includegraphics[width=0.9\textwidth]{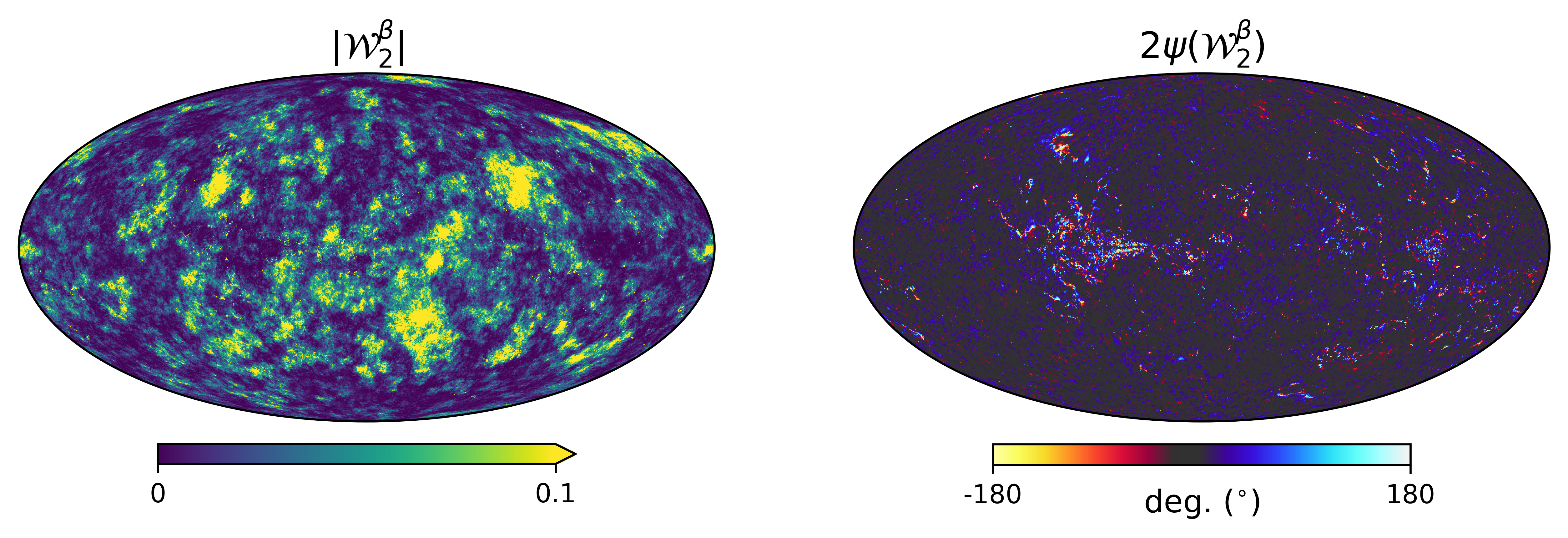}
    \caption{Maps of the modulus (left) and phase (right) of the complex spin moment $\mathcal{W}_2^\beta$ for the \dtwelve{} model.}
    \label{fig:W2b-d12}
\end{figure*}
\begin{figure*}[t!]
    \centering   \includegraphics[width=0.4\textwidth]{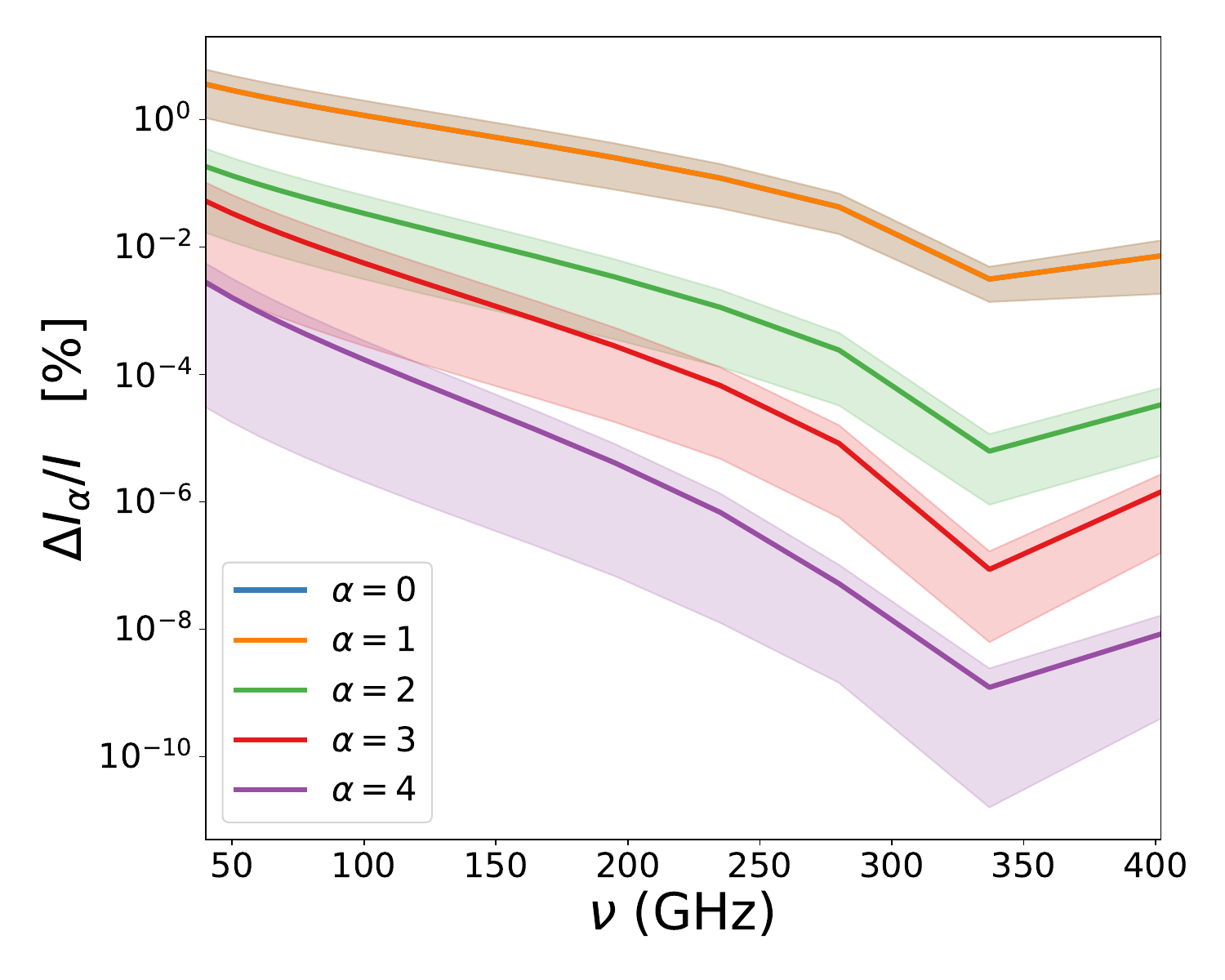}
    \includegraphics[width=0.4\textwidth]{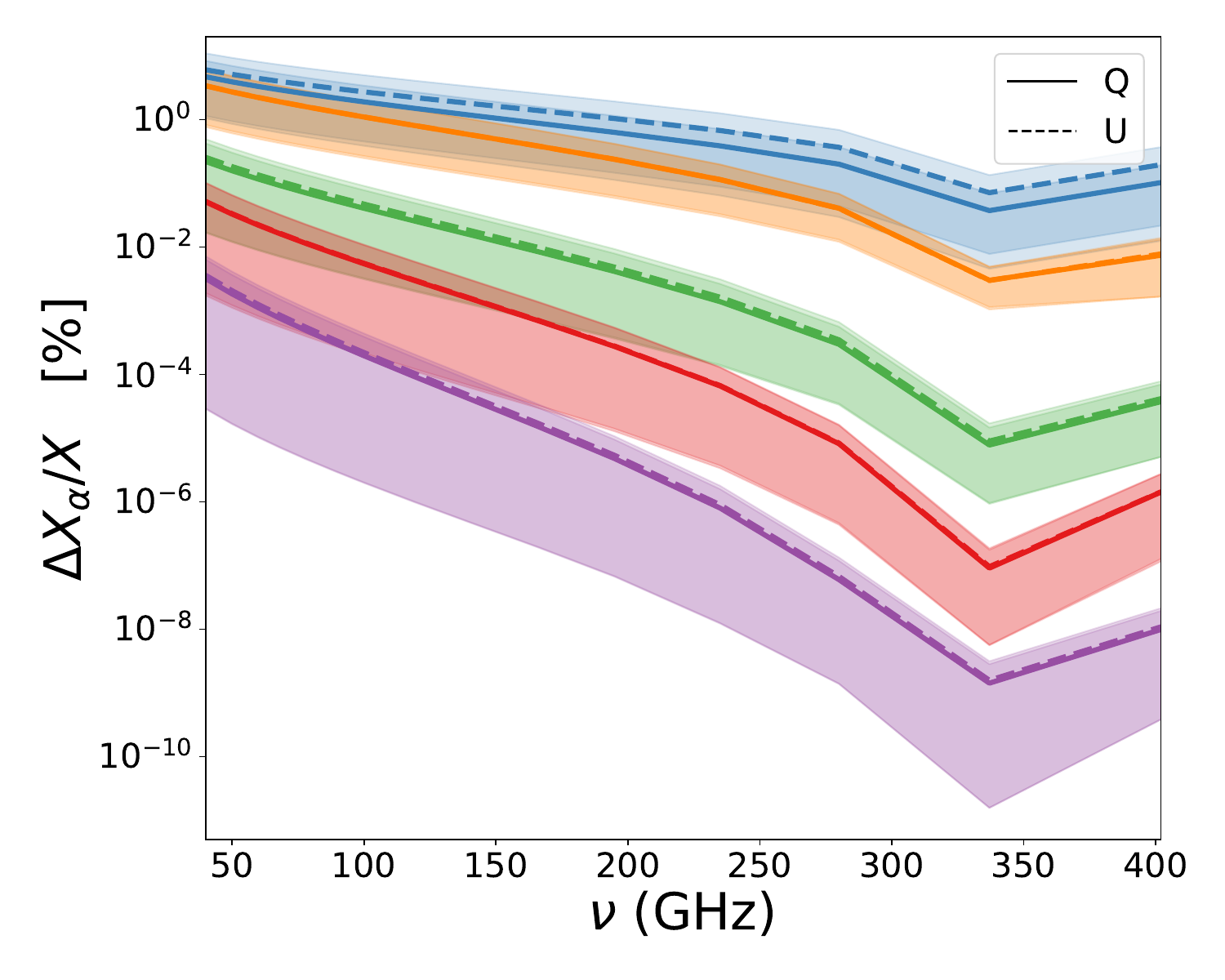}
    \caption{Residuals $\Delta X_\alpha/X$ of the spin-moment expansion for $\dtwelve{}$ expressed in percentages for the 15 bands of the $LiteBIRD$ instrument. In the left panel is $X=I$, and $X=Q,U$ is displayed in the right panel. The residuals were computed on the full sky up to a different order ($\alpha$) of the expansion. The lines correspond to the median values, while the shaded areas mark the values of the median absolute deviation. We note that the blue ($\alpha=0$) and the orange ($\alpha=1$) curves are superimposed in the left panel.}\label{fig:residuals}
\end{figure*}

\subsection{Thermal dust and modified blackbodies}

In this work, we focus on the thermal dust signal, which represents the main polarised foreground for CMB experiments at high frequencies ($\gtrapprox 70$\,GHz) \citep{Krachmalnicoff2016}. 
In the frequency interval relevant for CMB experiments, the SED of dust grains can be accurately modelled locally in both intensity and polarisation by the modified blackbody (MBB) function
\begin{equation}
    \mathcal{P}_\nu = \mathcal{A} \frac{B_\nu(T)}{B_{\nu_0}(T)}\left(\frac{\nu}{\nu_0}\right)^\beta,
\end{equation}
given by the product of a blackbody spectrum $B_\nu$ at temperature $T$ and a power-law with spectral index $\betad$, related to the grain properties. As such, the spectral parameters of this canonical SED are $\vec{p}=(\beta,T)$. The signal is polarised along the shape of the grain, such that $\psi={\rm Arg}(\mathcal{A})/2$ is perpendicularly aligned to the GMF, providing a powerful tracer of its structure \citep{Planck2016_XLIV}.

Due to the non-linearity of the MBB function, the total SED resulting from the mixing of multiple MBB with different spectral parameters will not be an MBB. It can, however, be accurately modelled using the moment expansion given in Eq.~\eqref{eq:mom-exp}, in which the derivatives are computed with respect to both $\betad$ and $T$. Because temperatures $T$ commonly have large values of $\mathcal{O}(10)$, the temperature differences involved in the moments can be significantly larger than 1. As such, the expansion in $T$ can diverge and one would prefer to perform an expansion with respect to the blackbody index $\tempd=1/T$ instead. 
Displaying the terms of the expansion only up to first order, the spin-moment expansion thus reads\footnote{Instead of dealing with complex quantities, one could perform two independent expansions in $Q$ and $U$ with different pivots ($\overline{\beta}^Q\neq \overline{\beta}^U$). Doing so reveals to be slightly less accurate than the spin-moment expansion but gives comparable results.}
\begin{align}
    \langle\mathcal{P}_\nu \rangle = \barP\times\Bigg( 1 + &\mathcal{W}_1^{\betad}\ln\left(\frac{\nu}{\nu_0}\right) +\mathcal{W}_1^{\tempd} \Delta\Theta_\nu+\cdots\Bigg),
\label{eq:mom-exp-MBB}
\end{align}
where $\Delta\Theta_\nu=\Theta_\nu-\Theta_{\nu_0}$ and $\Theta_\nu$ is the first order derivative of $B_\nu$ with respect to $\tempd$. The term $\barP$ is later referred to as the `overall MBB factor'. The exact expressions of the derivatives up to fourth order can be found for example in \cite{Vacher2022b}.

\section{A case study: Applying moments to dust models \label{sec:d12}}
\subsection{Modelling the signal in each pixel}

We now illustrate the power of the moment formalism both to model and understand the 3D complexity contained within a sky map. To do so, we use the foreground models available within the Python Sky Model (\texttt{PySM})  package\footnote{\url{https://github.com/galsci/pysm}} \citep{Thorne,Zonca2021}. \texttt{PySM} is a python software providing different types of data-driven full-sky simulations for the Galactic emissions. These simulations are largely used by the CMB community to apply and test component separation methods.
At present, the \dtwelve{} model is the only \texttt{PySM} dust model containing complexity in the third dimension along the line of sight. Using the terminology proposed in Sect.~\ref{sec:mom-rev}, \dtwelve{} is the only \texttt{PySM} model with intra-pixel complexity, that is with mixing occurring in each pixel at the native resolution of $N_{\rm side}=512$. At this resolution, the thermal dust signal in each pixel is generated as the superposition of up to six different layers of MBB emitting components as detailed in \cite{Martinez-Solaeche2018}. Each layer $l$ is associated with its own weight $\mathcal{A}_l$ and spectral parameters $\beta_l$ and $\tempd_l$. As it cannot be strictly given by an MBB, we would like to assess how well the signal of \dtwelve{} can be reproduced by a moment expansion performed in each pixel $\ipix$. To do so, we model the signal by the expansion displayed in Eq.~\eqref{eq:mom-exp-MBB}. The total complex amplitude is given by $\overline{\mathcal{A}}= \sum_l \mathcal{A}_l$\footnote{In agreement with the construction of \dtwelve{}, we chose the reference frequency $\nu_0$ to be 353\,GHz based on \planck{} highest frequency channel in polarisation. As such $\overline{\mathcal{A}}=\mathcal{P}_{\nu_0}$.}. The pivots $\overline{\betad}$ and $\overline{\tempd}$ can be obtained in every pixel $i_{\rm pix}$ by using Eq.~\eqref{eq:p_bar}\footnote{In order to avoid possible divergences of the pivot value outside the perturbative regime ($\sum_l{\mathcal{A}_l}\to 0$), we force the pivot to remain in the range $\overline{\betad} \in [0.9,2.5]$ and $\overline{\tempd} \in [1/35,1/10]$. If this criterion is not satisfied, we use the value of the intensity pivot to compute the spin moments. Further discussion on this point can be found in Appendix~\ref{app:normalisation}. \label{foot:pivot}}:
\begin{equation}
        \overline{\betad}(\ipix)= {\rm Re}\left(\frac{\sum_l \mathcal{A}_l \betad_l}{\sum_l\mathcal{A}_l}\right),
\qquad
        \overline{\tempd}(\ipix)= {\rm Re}\left(\frac{\sum_l \mathcal{A}_l \tempd_l}{\sum_l\mathcal{A}_l}\right).
\label{pbar-MBB}
\end{equation}
As further discussed in Sect.~\ref{sec:mom-rev}, this choice of pivot leads to the cancellation of the real parts of $\mathcal{W}_1^\betad$ and $\mathcal{W}_1^\tempd$, which ensures that the overall MBB factor is the closest possible MBB function we can use to model the \dtwelve{} signal in each pixel. To further clarify this statement, we considered a signal given by an exact MBB with parameters $\beta$ and $\tempd$. Trying to model this signal with a moment expansion using pivots $\beta'\neq \beta$ and $\tempd'\neq \tempd$ would lead to a non-zero value for all the moments including the first first orders. However, it would be incorrect to interpret these moments in term of deviations of the signal from an MBB function, as they simply capture the difference between the spectral parameters of the signal and the one of the overall MBB factor used in the expansion. Using instead the adjusted pivot as done here would lead to a cancellation of all the moments, revealing that the signal is indeed a perfect MBB. Any remaining non-zero moments above first order, however, would indicate deviations of this signal that cannot be captured by an MBB function.
As such, with our choice of pivot, we ensure that the moment terms model only deviation to the MBB function (i.e. polarised SED distortions and rotation of the polarisation angle with frequency). The first order moments are purely imaginary and expected to contribute significantly to the spectral rotation of the polarisation angle. 

We also emphasise that we computed a value of $\overline{\betad}$ and $\overline{\tempd}$ in each pixel. As such, the 2D (inter-pixel) variability of the model is expected to be entirely contained within the overall MBB factor, while the moment coefficients grasps only the intra-pixel complexity interpreted as extra variations along the third dimension. Another valid approach would have been to consider a single value of $\overline{\betad}$ and $\overline{\tempd}$ for the whole sky. In this case, the moment coefficients would have accounted for both the 2D (inter-pixel) and the 3D (pixel) complexity indistinguishably. 

The spin-moment coefficients of the equations can be computed from the layers using the discrete generalisation of Eq.~\eqref{eq:moments} as
\begin{equation}
\mathcal{W}_\alpha^{\betad^\delta \tempd^\gamma}(\ipix)= \frac{\sum_l \mathcal{A}_l (\betad_l-\overline{\betad}(\ipix))^\delta(\tempd_l-\overline{\tempd}(\ipix))^\gamma}{\sum_l\mathcal{A}_l},
\label{eq:mom-d12}
\end{equation}
where $\alpha=\delta+\gamma$. As such, expanding up to first order, one adds to the MBB the two purely imaginary moments $\mathcal{W}_1^\betad$ and $\mathcal{W}_1^\tempd$. Up to second order, the three complex moments $\mathcal{W}_2^\betad$, $\mathcal{W}_2^\tempd$ and $\mathcal{W}_2^{\beta \tempd}$ are added to the expansion and the third order sees the inclusion of four new complex moment maps $\mathcal{W}_3^\betad$, $\mathcal{W}_3^\tempd$, $\mathcal{W}_3^{\betad^2 \tempd}$ and $\mathcal{W}_3^{\betad \tempd^2}$ etc.

\review{The moment maps are quantifying the complexity contained in the third dimension.}
\review{As complex numbers, they can be characterised either by their real and imaginary parts or by their modulus and phase.}\reviewcorrect{As illustrated in Fig.1 with $\mathcal{W}_2^\betad$, the moment maps are a proxy of the complexity in the third dimension.}{}\review{As an illustration, the modulus and phase of $\mathcal{W}_2^\betad$ are displayed in Fig.~\ref{fig:W2b-d12}.} \reviewcorrect{For example,}{Being the second order moment,} $\mathcal{W}_2^\beta$ informs us about the \reviewcorrect{standard deviation}{variance} of the $\betad$ distribution along each line of sight, weighted by a contribution from the \reviewcorrect{angle distribution}{distribution of the polarisation angles\footnote{\review{Looking at Eq.~\ref{eq:mom-d12}, we can identify $\mathcal{W}_\alpha^{\beta^\delta \mathcal{T}^\gamma}$ as the statistical moments (variance, skewness, kurtosis ...) of order $\alpha$ of the discrete $\beta-\mathcal{T}$ distribution where the $\mathcal{A}_l$ (of phase $2\psi_l$) are complex weights associated to each pair of $(\beta_l,\mathcal{T}_l)$.}}}. \review{The modulus $|\mathcal{W}_2^\beta|$ can be directly compared with the variance of the $\beta$ distribution contained in each pixel, on the left lower panel of Fig.~\ref{fig:var2D-3D-d12}. From the map of $|\mathcal{W}_2^\beta|$, we understand that the intra-pixel complexity of the \dtwelve{} model is distributed over the sky as a Gaussian random field with clusters of higher complexity at intermediate latitudes. We thus expect the effects of mixing to be more important in the bright regions, claim which can be verified by comparing this map with the map of the distortions of the polarised SED, which are represented on the upper left panel of Fig.~\ref{fig:momd12-correl}.} On the other hand, \reviewcorrect{its}{the} phase \review{map} $2\psi(\mathcal{W}_2^\beta)$ probes more directly regions of high $\psi$ complexity, which correlate with the circular variance of the $\psi_l$ in each pixel represented on the right lower panel of Fig.~\ref{fig:var2D-3D-d12}. Without any variations of $\psi_l$ (i.e. of the magnetic field), this phase should be strictly zero, and the moments purely real quantities\footnote{This can be seen by considering Eq.~\eqref{eq:mom-d12} with $\mathcal{A}_l=A_le^{2\psi_0}$ where $\psi_0$ is a constant polarisation angle shared by all the layers.}. Intuitively, the phase \review{of the moment maps} represents the ability of the moment to capture the \review{frequency} rotation of the original signal in the complex plane. \review{One can see the strong correlation between this phase map with the upper right panel of Fig.~\ref{fig:momd12-correl}, representing the rotation of the polarisation angle in each pixel of the \dtwelve{} model.} \reviewcorrect{From Fig.~1, one can observe that the intra-pixel complexity of \dtwelve{} is distributed on the map as a Gaussian random field with clusters of higher complexity at intermediate latitudes.}{}
\review{All the moment maps involved for the second order expansion are displayed in Figs.~\ref{fig:allmom-d12-modphase} and \ref{fig:allmom-d12}, and w}e comment further on \reviewcorrect{the values of the moment maps}{their values} and their interpretation in Appendix~\ref{app:mom-maps}. 
\review{We also note that the moment maps thus defined probe only the inter-pixel/3D properties of the models and are independent of the inter-pixel complexity of \dtwelve{} that is of its structure on the sky. This property is proven in Appendix~\ref{app:universality} and can also be verified by comparing the structure of the moment maps with the structure of the \dtwelve{} model at $\nu_0$ displayed on the upper panels of Fig.~\ref{fig:var2D-3D-d12}.}

In order to evaluate the accuracy at which the moment expansion can reproduce the {\tt d12} dust model, we considered the residuals defined as
\begin{equation}
    \frac{\Delta X_\alpha}{X}= \left|\frac{X^{\tt d12}-X^{\rm mom}_{\alpha}}{X^{\tt d12}}\right|,
\end{equation}
with $X \in [I,Q,U]$, and $X^{\rm mom}_{\alpha}$ is the value of $X$ estimated using the moment expansion Eq.~\eqref{eq:mom-exp-MBB}, including all the terms up to order $\alpha$.  
In principle, such residuals can be computed at any frequency $\nu$. In order to provide an illustration on a broad frequency interval relevant for CMB missions, we consider the fifteen bands of the $LiteBIRD$ instrument between 40 and 402 GHz \citep{Ptep}.
The median value and median absolute deviation of the residuals are displayed in Fig.~\ref{fig:residuals} for the LiteBIRD frequency range. Statistics are computed considering all the pixels on the full sky without masking and the residuals are displayed for a moment expansion performed up to different orders between zero and four. For a more quantitative estimation, the corresponding median values are reported in Tab.~\ref{tab:med_val} for $\nu=100$ GHz, corresponding to a frequency band significantly different from the reference frequency ($\nu_0=353$ GHz) where the dust signal is still dominant.

\begin{table}[]
    \centering
    \begin{tabular}{c}
    {$\Delta X_{\alpha} / X$}
    \end{tabular} \\
    \begin{tabular}{c|ccccc}
          $X$ & 
          $\alpha=0$ & $\alpha=1$ & $\alpha=2$ &$\alpha=3$ & $\alpha=4$\\
          \hline
          $I$ & 1.16 \% & 1.16 \% & 0.034 \% & 0.0056 \% & 0.00017 \% \\
          $Q$ & 1.90 \% & 1.10 \% & 0.041 \%& 0.0055 \% & 0.00020 \%\\
          $U$ & 2.73 \% & 1.10 \% & 0.047 \% & 0.0055 \% & 0.00022 \%
    \end{tabular}
    \caption{Median value of the residual distribution of Fig.~\ref{fig:residuals} estimated at $\nu=100$ GHz.}
    \label{tab:med_val}
\end{table}

One can notice clearly that the moment expansion significantly lowers the residuals by around one order of magnitude from one order to the other, thus improving the accuracy of the modelling of the intra-pixel complexity as expected\footnote{We find that an expansion in $\tempd$ is $\sim 20\%$ more accurate than an expansion in $T$ in both $Q$ and $U$ at order $\alpha=2$, justifying a posteriori our choice to use $\tempd$ instead of $T$.}. 
Looking now at each individual order:
\begin{itemize}
    \item The zeroth order is given by an MBB function with a different $\overline{\betad}$ and $\overline{\tempd}$ in each pixel. With our choice of pivots, we expect this model to provide the closest possible modelling of \dtwelve{} with MBB SEDs. We see that for the three states $(I,Q,U)$, the MBB already reproduces the $\dtwelve{}$ model at the percent level, revealing that the \dtwelve{} model does not inherit a lot of complexity from its third dimension. We can notice that the modelling is around two times worse for polarisation than for intensity, hinting for higher inherent complexity of the polarised signal.  
    \item The addition of the first order terms does not impact the intensity residual. This is expected as our choice of pivot values is designed to cancel any first order contribution. This is not the case for polarisation, where the imaginary contribution remains for the first order moments. The addition of this imaginary contribution drives the residuals of polarisation down to the level of the intensity ones. This reflects the fact that any polarisation angle rotation with frequency (i.e. a difference of SED between $Q$ and $U$) could not be captured by the zeroth-order.
    \item The addition of higher order terms significantly lowers the value of residuals both for intensity and polarisation. At second order of the expansion, \dtwelve{} is recovered up to an accuracy of a few per ten thousand $(0.01\%)$. Pushing the expansion up to fourth order, we reach a per million $(10^{-4}\%)$ accuracy level for the three states $I$, $Q$ and $U$.
\end{itemize}
As such, moment coefficients clearly provide a compact and well motivated way to quantify and model any complexity (whether inter-pixel or intra-pixel) arising from mixing. 
Keeping all the moment contributions, one needs $2\alpha^2+4$ parameters to model the signal at order $\alpha$\footnote{Including $\mathcal{A}, \overline{\betad}, \overline{\tempd}$ and disregarding the real part of the first order moments which have been purposefully cancelled.}, while one needs $4\times \mathcal{N}_l$ parameters to describe the sum of $ \mathcal{N}_l$ modified blackbodies\footnote{For each layer: two parameters for $\mathcal{A}_l=Q_l+ \i U_l$ as well as $\beta_l$ and $T_l$}. Thus in the present situation, we used 12 parameters at second order to model the impact of up to 24 parameters for the six layers of \dtwelve{}, reproducing the original signal with an accuracy of a few per ten thousands with two times less parameters. Finally, we note that the moment expansion, as a Taylor expansion, could also be able to account for more general deviation to the MBB emission-law having different origin than mixing, but this point remains to be explicitly addressed in the literature. Moreover, moment coefficients are
agnostic on the number of layers and/or the specific values along the line of sight. Indeed, what is commonly modelled as discrete sums should be integrals over continuous distributions, which can then still be accurately captured with the moment expansion in an economic way.

\subsection{The proxies of complexity in harmonic space}

We now move into harmonic space and investigate two proxies of the complexity of the foreground signal: the decorrelation and the $E/B$ ratio of the model.
As its name suggests, decorrelation quantifies the loss of correlation between two maps at frequencies $\nu_i$ and $\nu_j$ due to mixing \citep{Planck2016_XXX}. Decorrelation can be quantified in harmonic space by the deviation from unity of the so-called spectral correlation ratio. Focusing only on the $B$-mode signal, it is defined as 
\begin{equation}
\Delta^{BB}_\ell(\nu_i\times\nu_j) = \frac{\mathcal{C}^{BB}_\ell(\nu_i\times\nu_j)}{\sqrt{\mathcal{C}^{BB}_\ell(\nu_i\times\nu_i)\mathcal{C}^{BB}_\ell(\nu_j\times\nu_j)}}.
\end{equation}

Another proxy of complexity is given by the frequency dependence of the $E/B$ ratio, which is a consequence of mixing directly related to the co-variation of polarisation angles and spectral parameters.\footnote{This effect has been discussed in length in \cite{Vacher2023} and some possible evidences in \planck{} data were be found in \cite{Ritacco22}.}
We chose to look at the frequency dependence of the quantity
\begin{equation}
r_\nu^{E/B} = \frac{\langle\mathcal{C}_\ell^{EE}\rangle_\ell}{\langle\mathcal{C}_\ell^{BB} \rangle_\ell}(\nu),
\end{equation}
where the average $\langle ... \rangle_\ell$ is done over all multipole $\ell$-bins considered in the analysis.

Next, we intend to compare these two quantities for the maps of \dtwelve{} and its moment expansion at different orders. To do so, we use the \planck{} Galactic mask\footnote{\url{http://pla.esac.esa.int}.} with $f_{\rm sky}=70\%$ apodised with a $5^\circ$ degree scale. All maps are computed with a resolution of $N_{\rm side}=512$ and at the four highest frequency bands of \planck{}, $\nu=[100,143,217,353]$ GHz.
Angular power-spectra are computed using {\sc Namaster}\footnote{\url{https://github.com/LSSTDESC/NaMaster}} with the $B$-modes purification \citep{namaster}. All band-powers are binned 10 by 10 from $\ell=2$ up to $\ell=2N_{\rm side}+1$.
\begin{figure*}[t!]
    \centering    \includegraphics[width=\columnwidth]{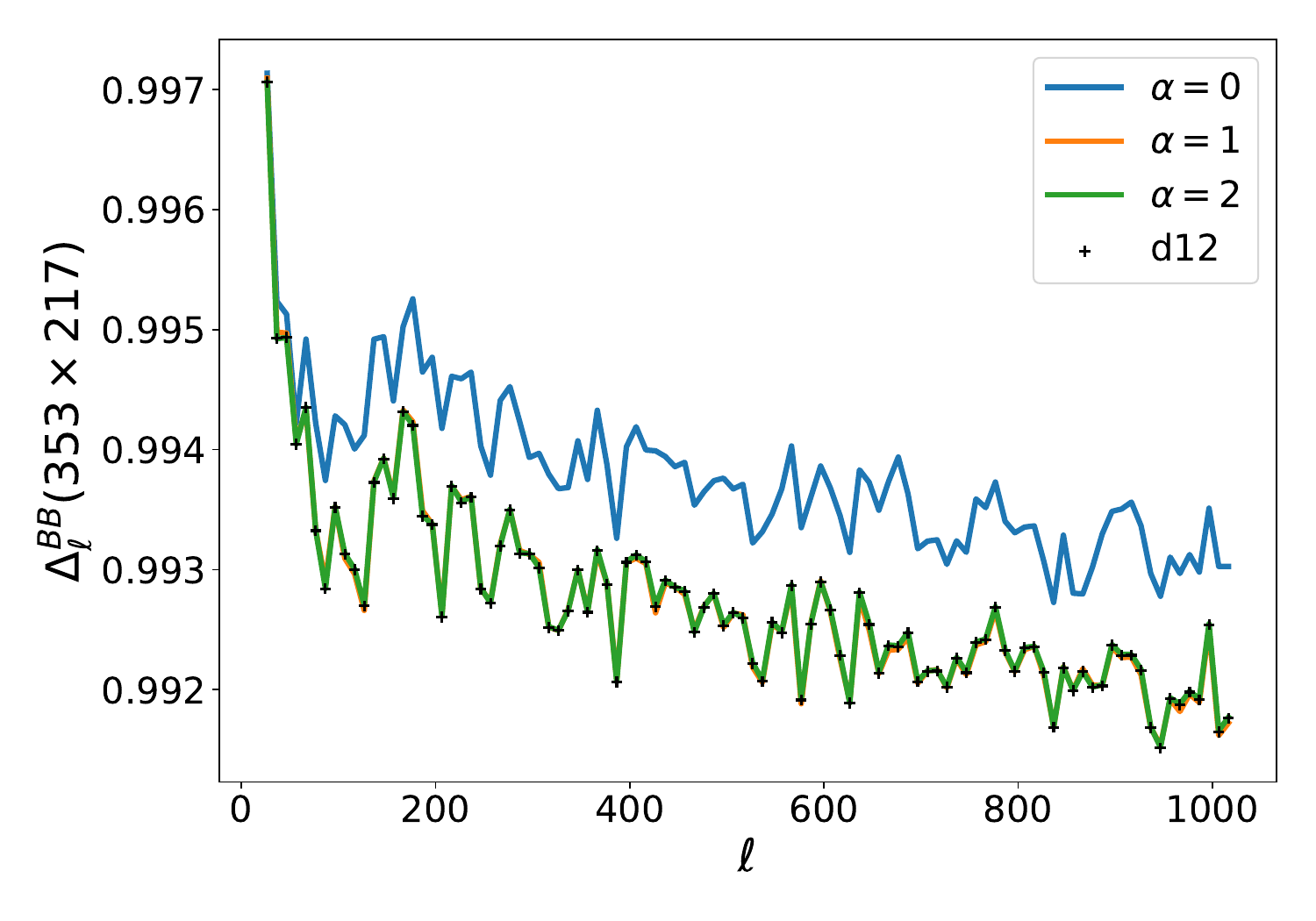} \includegraphics[width=\columnwidth]{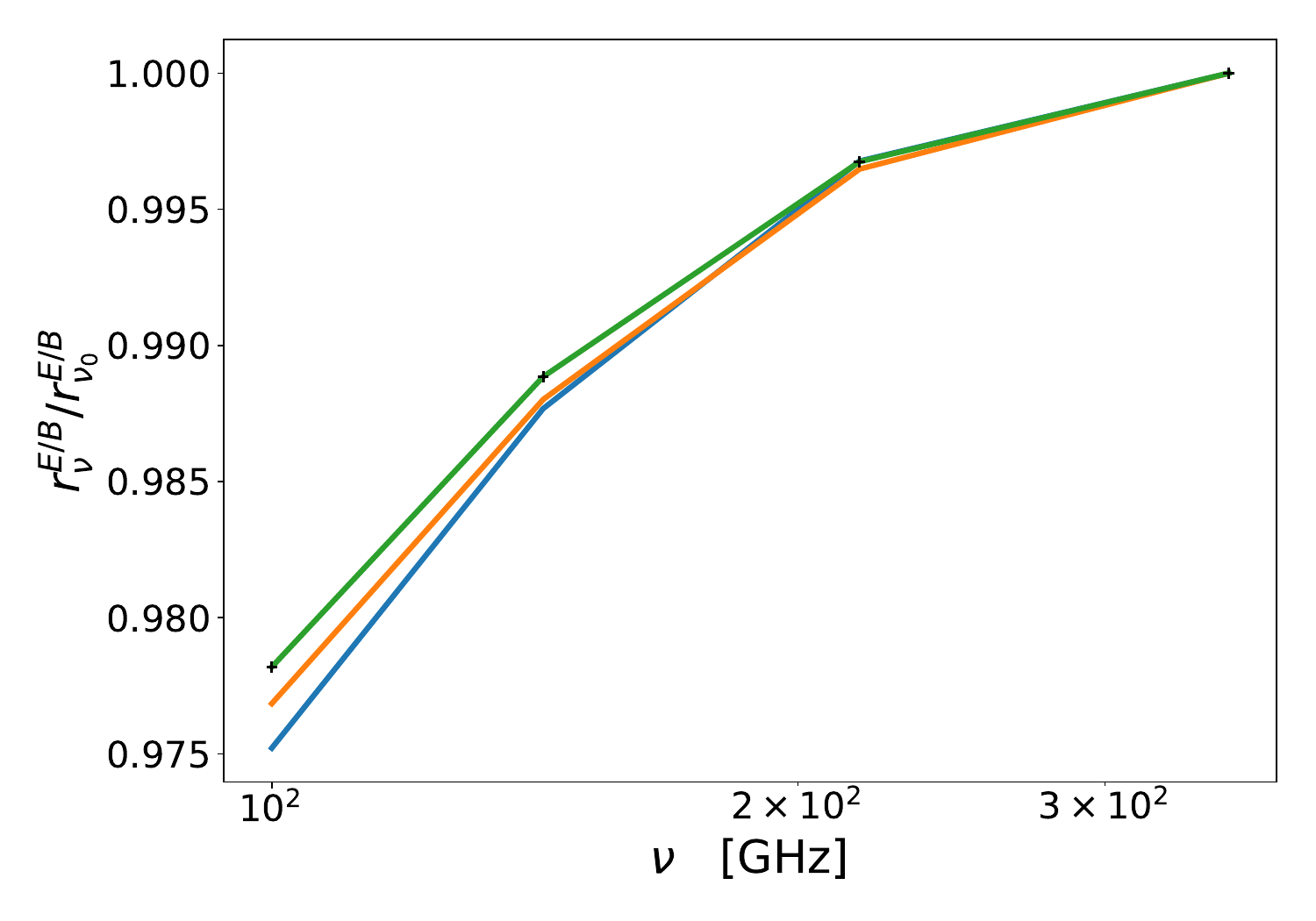} 
    \caption{Two proxies of the dust complexity: decorrelation $\Delta^{BB}_\ell(353\times 217)$ (left) and the $E$-to-$B$ ratio $r^{E/B}$ (right). Values are plotted for the \dtwelve{} original map (black crosses) and its representation by a pivot modified blackbody (blue), a moment expansion up to first order (orange) and up to second order (green).}
    \label{fig:Delta-reB-d12}
\end{figure*}

The results can be found in Fig.~\ref{fig:Delta-reB-d12}. We observed that the expansion performed up to the second order allow both proxies to be modelled with an extremely high accuracy. The first order expansion almost enables the gap of decorrelation coming from the third dimension to be fully filled, while the expansion has to be performed up to the second order to catch the variations of $r^{E/B}_\nu$. As illustrated in Appendix~\ref{app:mom-maps}, this last fact can be understood by looking at the difference between the $EE$ and the $BB$ spectra of the moment maps, which is large when computed for the maps of $\mathcal{W}_2^\beta$. This illustration shows how powerful the moment maps can be at quantifying and modelling the complexity of a dust model within a few sets of interpretable coefficients. In theory, any foreground model, whether dust or synchrotron, can be fully represented by a set of moment maps.\footnote{In the simpler case where no intra-pixel complexity is included, one can use a single constant pivot over the full sky, reducing even further the number of parameters used.}
Regarding both proxies and comparing \dtwelve{} with the MBB case ($\alpha=0$), for which the signal is given by $\barP$ without moments, we see that the overall MBB factor -- capturing all the information about the inter-pixel complexity -- represents the biggest contribution to the decorrelation and the $E/B$ ratio variation of \dtwelve{}. We can thus conclude that \dtwelve{} is indeed relatively complex, but that this complexity arises mostly from the spatial variation of the emission properties in 2D (inter-pixel) and not so much from variations within the third dimension (intra-pixel). From this result, one can thus wonder how much complexity can be injected in the third dimension of a model while remaining compatible with data and what could be the impact of such a complexity.

\section{Exploring alternative models for the complexity in the third dimension using moments \label{sec:model-building}}

Next, we use the moments and their ability to model complexity in order to build new simulations of the dust emission. Our goal is to mimic mixing occurring along the line of sight with the addition of intra-pixel complexity with a set of moment maps. Obviously, the most reasonable way to proceed would be to infer this information from observational data. Unfortunately, such a complete set of information would require accurate 3D mapping of the full sky values of $\betad$, $\tempd$, $I$, $Q$, $U$\footnote{Alternatively and more physically, one would need to trace the dust column density and the angle of the GMF in the three-dimensional Galaxy.}, which is not yet available [some works in that direction have been undertaken recently: a map of the temperature on $3\pi$ can be found in \cite{Zelko2022} and a portion of the magnetic field has been mapped by \cite{Pelgrims2024}].

While we currently lack such observationally driven moment maps over the full sky, we can investigate the impact of any arbitrary moment map introducing some additional complexity in the third dimension. In particular, we can assess the impact of the addition of moments on various observables, as the value of the angular power spectra or the decorrelation, and infer from these observables the maximal amplitude of the moments allowed by the current limit set by the measurements of the \planck{} mission. 
A simple way to move forward in this direction is to re-use the set of moment maps we already have at hand from the \dtwelve{} analysis and re-inject them in a simpler model. By multiplying the moment templates by constant factors, we will be able to assess their impact on different observables and fine-tune a model containing the maximal amount of complexity with regard to limits set by current data. The goal is then to see if and how this additional 3D complexity impacts component separation for future CMB missions.
As further discussed in Appendix~\ref{app:mom-maps}, we can strongly doubt that the \dtwelve{} moments faithfully describe the dust complexity along the third dimension/line of sight. Specifically, they do not account for the non-Gaussian properties of the ISM. Nevertheless, the fact that they were extracted from some credible realisations of a distribution of $\betad$ and $\tempd$ suggests the absence of any too pathological behaviour.\footnote{However, we that infinitely many different realisations of $\beta$ and $\tempd$ could produce identical moment maps.} They thus provide a reasonable playground to investigate the addition of complexity at the pixel level. 

As most of the complexity of \dtwelve{} is arising from variations of the spectral parameters in the plane of the sky (inter-pixel mixing), we suggest to consider instead the milder \texttt{PySM} \dten{} model as our baseline. This model, based on a refined GNILC analysis of the \planck{} data \citep{2016A&A...596A.109P,Planck18_XII} is commonly considered as a reasonable reference by the CMB community \citep{Wolz2024,Spider2025}. On the other hand, the maps of layers of $\beta$ and $\tempd$ of \dtwelve{} were constructed as Gaussian random realisations around the best fit values obtained in \cite{Planck2014dust}. Contrarily to \dtwelve{}, \dten{} assumes a perfect MBB scaling in every pixel of the sky (at $N_{side}=512$ resolution) thus  containing only inter-pixel complexity arising from the variation of both $\betad$ and $\tempd$.  
We intend to inject further complexity in each pixel using \dtwelve{}'s moment maps and quantify the impact of such addition. As discussed in Appendix \ref{app:universality}, thanks to our choice of normalisation and pivot for the moments, the \dtwelve{} moment maps contain only information about the variations along the line-of-sight but are completely agnostic on the specific values of the pivot $\bar{p}$ and amplitude $\mathcal{A}$ of \dtwelve{}. As such, they can be applied to any other map without introducing additional spurious and unwanted complexity from the inter-pixel properties of \dtwelve{}.

To simplify our discussion, we limit ourselves to the inclusion of moment contributions up to second order as we found in the previous section that this was enough to reproduce the signal of the model with an accuracy reaching the one in ten thousand level as well as the two harmonic proxies of complexity $r^{E/B}_\ell$ and $\Delta^{BB}_\ell$.
\begin{figure*}[t!]
    \centering    \includegraphics[width=\columnwidth]{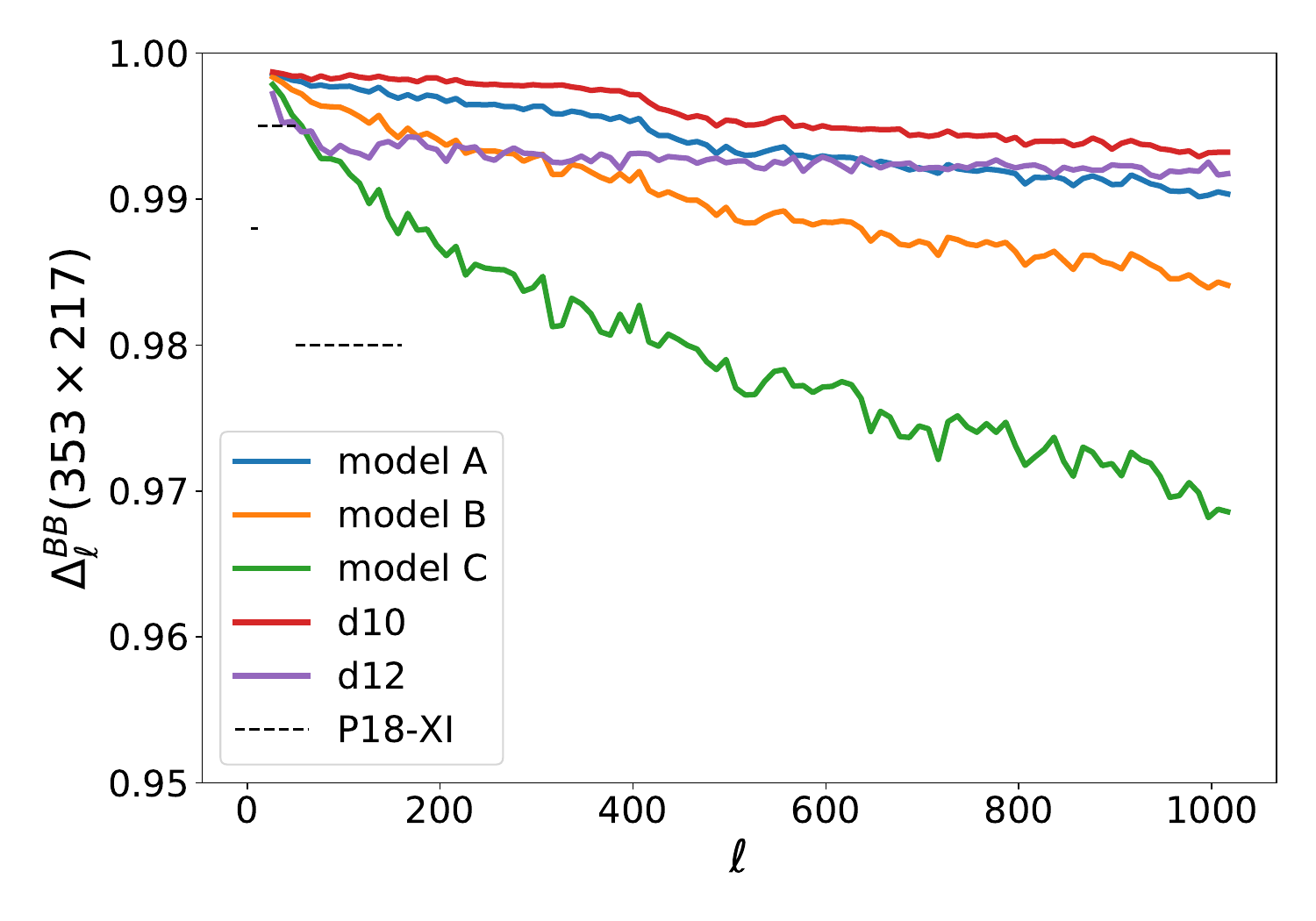} \includegraphics[width=\columnwidth]{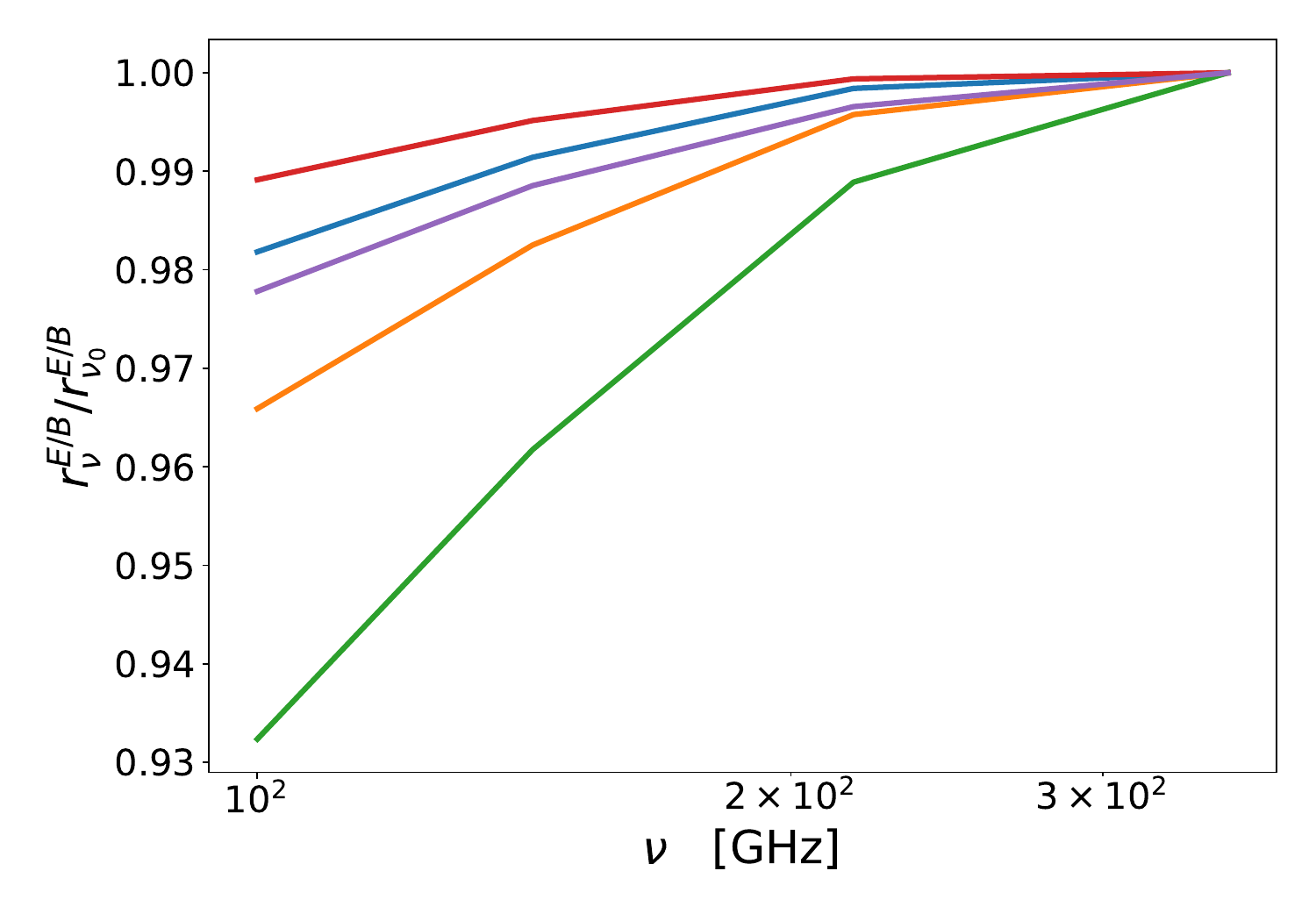}
    \caption{Comparison of the proxies of complexity, $\Delta_\ell^{BB}$ (left) and $r^{E/ B}_\nu$ (right), for the three proposed models (A, B, and C), \dten{} and \dtwelve{}. The black dashed lines in the left panel represent the lower limits on $\Delta_\ell^{BB}$ set by the analysis of \planck{} data and taken from \cite{Planck18_XI}.}
    \label{fig:model-comp-Delta-rEB}
\end{figure*}

\subsection{Model building}

We propose to create models including complexity both in the plane of the sky (inter-pixel level) and along the line of sight (intra-pixel level) using Eq.~\eqref{eq:mom-exp-MBB}. The inter-pixel complexity is contained within the sky varying modified blackbody $\barP$ of \dten{}, where $\bar{
\betad}(\ipix)$ and $\overline{\tempd}(\ipix)$ are maps containing variations over the sky. The intra-pixel complexity on the other hand will be given by the moment maps $\Walphap(\ipix)$ derived from the intra-pixel complexity of the \dtwelve{} model. 

We consider the three following models:
\begin{itemize}
    \item \textbf{Model A:} We first consider the model given by \dten{} to which complexity is added using exactly the \dtwelve{} moments through Eq.~\eqref{eq:mom-exp-MBB} up to second order in each $N_{\rm side}=512$ native pixel. This model is expected to be less complex than \dtwelve{} as it does not inherit from its 2D-complexity and thus contains less inter-pixel mixing, but it will surely be more complex than \dten{} due to the addition of intra-pixel complexity with the moments.
    \item \textbf{Model B:} To build this model, we start from model A and multiply all the moment maps by a unique constant factor. 
    Doing so is equivalent to rigidly amplify all the statistical moments (except the mean) of the distributions of $\beta$ and $\tempd$ as well as the variation of the angles along the line-of-sight.
    \item \textbf{Model C:} In this case, only the imaginary part of the moments of model A are amplified by a constant factor. As detailed in Appendix~\ref{app:mom-maps}, doing so should enhance the frequency dependence of the polarisation angles. As shown in Fig.~\ref{fig:Delta-reB-d12}, the first order (which is purely imaginary) is the main driver of the decorrelation for \dtwelve{}. Therefore, enhancing only the imaginary part is expected to increase $\Delta_\ell^{BB}$ significantly.
\end{itemize}
The value of the constant factor for each model ($x_B=1.8$ and $x_C=3$ for models B and C respectively) is chosen such that it takes the maximal value allowed by the \planck{} data. Each model is indeed confronted with the values of the dust angular auto-power spectra measured by the \planck{} mission in Sect.~\ref{sec:comp-Planck}. Then, we look at the proxies of the complexity $\Delta^{BB}_\ell$ and $r^{E/B}_\nu$ for each model in Sect.~\ref{sec:proxies-compl}. An interpretation of the values of the proxies in terms of moment maps is also provided. Finally,  in Sect.~\ref{sec:compsep} we address the subtraction of the simulated dust emission through standard component separation pipelines to assess the impact of the added three-dimensional complexity on CMB observations.

\subsection{Comparison with \planck{} data \label{sec:comp-Planck}}

We now compare the three models A, B and C with the measurements of the polarised angular power-spectra by the \planck{} mission. \review{Our goal is to use the  measurements and error-bars of \planck{} in order to set an upper limit on the complexity that can be added to the new dust models.} To do so, we defined
\begin{equation}
\delta \mathcal{C}_\ell^{XX} = \frac{|\mathcal{C}_\ell^{XX}(\dten{})-\mathcal{C}_\ell^{XX}({\rm model})|}{\sigma_{\rm Planck}}
\label{eq:deltacl}
\end{equation}
as the distance of our model to \dten{} in units of the \planck{} noise $\sigma_{\rm Planck}$. Here, $XX\in\{EE,BB\}$ and $\sigma_{\rm Planck}$ is an estimation of the \planck{} noise computed from the standard deviation of 600 \texttt{NPIPE} simulations of the sky as observed by \planck{} \citep{NPIPE}. The details of the derivation of $\sigma_{\rm Planck}$ can be found in Appendix~\ref{app:error-bars}.

In order to be conservative, we consider that a model is rejected by data if $\delta \mathcal{C}_\ell^{XX}> 1$. The value of  $\delta \mathcal{C}_\ell^{EE}$ and $\delta \mathcal{C}_\ell^{BB}$ for the three models can be found in Appendix ~\ref{app:power-spectra} for the 10 cross-spectra associated to the four highest frequencies of \planck{} in polarisation. For $\nu=\nu_0=353$\,GHz, we verify that $\delta \mathcal{C}_\ell^{XX}=0$, for all the models, which must be true by construction. 

We see that model A produces a change in the angular power spectra which is well below the limit of \planck{} data. Model B is instead created by multiplying all the moments of the model A by a constant number $x_B$ until $\delta \mathcal{C}_\ell^{XX}$ reaches the limit of $\delta \mathcal{C}_\ell^{XX}=1$. We find that this limit is set by $\delta \mathcal{C}_\ell^{EE}(143\times 217)$ to be $x_B=1.8$. The model C is built by multiplication of only the imaginary part of the moments by a constant number $x_C$. We chose here $x_C=3$ and the model does not reach the limit of $\delta \mathcal{C}_\ell^{XX}=1$ for any combination of \planck{} frequencies. However, values of $x_C>3$ are rejected by the limit set by \planck{} data on the decorrelation (left panel of Fig.~\ref{fig:model-comp-Delta-rEB}), which be discussed in the next section.

\subsection{Proxies of complexity \label{sec:proxies-compl}}

For all the models, we computed $\Delta_\ell^{BB}$ and $r^{E/B}_\nu$ on 70\% of the sky as detailed in Sect.~\ref{sec:d12}. The results can be found in Fig.~\ref{fig:model-comp-Delta-rEB}. Lower limits on $\Delta_\ell^{BB}$ have been added in black dashed lines, taken from \cite{Planck18_XI} (Appendix B, Table B.1, mask `LR71'). We note that these limits are only available at rather large angular scales, which fortunately are also the most relevant to $B$-mode searches.

Regarding both proxies, the model A is only slightly more complex than \dten{} and less complex than \dtwelve{}. Such a result is expected as most of the complexity from \dtwelve{} comes from the variations of its parameters on the plane of the sky as detailed in Sect.~\ref{sec:d12}. However, at the smallest scales ($\ell\geq 500$), the decorrelation of the model A becomes comparable to \dtwelve{}. 

Model B has a larger decorrelation than \dtwelve{} for $\ell\geq 200$ and a larger $r^{E/B}_\nu$ for all frequencies. We thus expect this model to have either a stronger or a comparable impact than \dtwelve{} on component separation and $B$-mode recovery.

Model C is by far the most complex model regarding both proxies. It reaches the limits of decorrelation set by \planck{} data for $\ell\in[11,50]$ and has a greater $\Delta_\ell^{BB}$ than \dtwelve{} for $\ell \geq 50$. Its value of $r^{E/B}_\nu$ is greater than \dtwelve{}'s for all frequencies. Among the considered models, we thus expect this scenario to have the maximal impact on component separation and $B$-mode science.

\subsection{Component separation \label{sec:compsep}}

We now want to study the impact of the addition of pixel level complexity on component separation methods for future CMB missions. We consider a 'LiteBIRD-like' experiment according to the instrumental specifications reported in \citet{Ptep}. In order to isolate the impact of the various dust models, in this analysis we consider the simplest synchrotron model available in the literature, the \texttt{PySM} \texttt{s0}, which assumes a power-law SED with isotropic spectral index $\beta_s$ across the sky and against which the two considered component separation methods have already proven to be robust. At each frequency of the data set, the foreground signal therefore includes simulated synchrotron and dust maps, with the latter varying depending on the considered model. To them, multiple realisations of CMB and instrumental noise are coadded. CMB maps are simulated through the \texttt{HEALPix} package \citep{Gorski2005} from the best-fit theoretical angular power spectrum of $2018$ \planck{} data  \citep{2020A&A...641A...6P}. Instrumental noise in each pixel is a random realisation drawn from a Gaussian distribution with standard deviation associated to the anticipated sensitivity of the corresponding frequency map \citep{Ptep}. We consider the application of two component separation approaches: parametric and internal linear combination (ILC) \citep{ILC}, to highlight how they can be differently affected by the complexity of the dust emission. We applied simple incarnations of the two methodologies: the multiresolution FGBuster pipeline \citep{FGBuster,Errard2019} with spectral parameters fitted in different patches across the sky and the needlet ILC (NILC, \cite{NILC}). 

The FGBuster pipeline assumes a parametric model for the SEDs of the different components of the Galactic emission and fit it to the data. Specifically, best-fit values of the spectral parameters are derived by maximising the spectral likelihood defined in \citet{FGBuster}. A different value of a spectral parameter can be fitted in each pixel of an \texttt{HEALPix} grid with predefined resolution parameter $N^{\textrm{FGB}}_{\textrm{side}}$. In this analysis, we assume a power-law and an MBB as sycnhrotron and dust model, respectively. Through the FGBuster pipeline, we fit different values of the dust spectral index $\betad$, and temperature $T$ in \texttt{HEALPix} patches with $N^{\textrm{FGB}}_{\textrm{side}}=64$ for $\betad$ and $N^{\textrm{FGB}}_{\textrm{side}}=[8,4,0]$ for $T$, where the three values are associated to different regions defined by the \planck{} HFI masks as proposed in \cite{Ptep}. Only a single value is fitted for the spectral index of synchrotron $\beta_s$, as no anisotropies of synchrotron spectral properties are injected in the simulations. 
The NILC technique \citep{NILC}, instead, combines data at different frequencies with frequency- and pixel-dependent weights so as to minimise locally the variance of the output map while preserving the full blackbody signal. In order to properly address the different contamination from foreground and noise, the pipeline is applied separately at different angular scales making use of the needlet filtering \citep{2006math......6599B}. In this analysis, we adopt the standard needlet construction \citep{stand_nls}. 

\begin{figure*}[t!]
    \centering 
    \includegraphics[width=0.97\columnwidth]{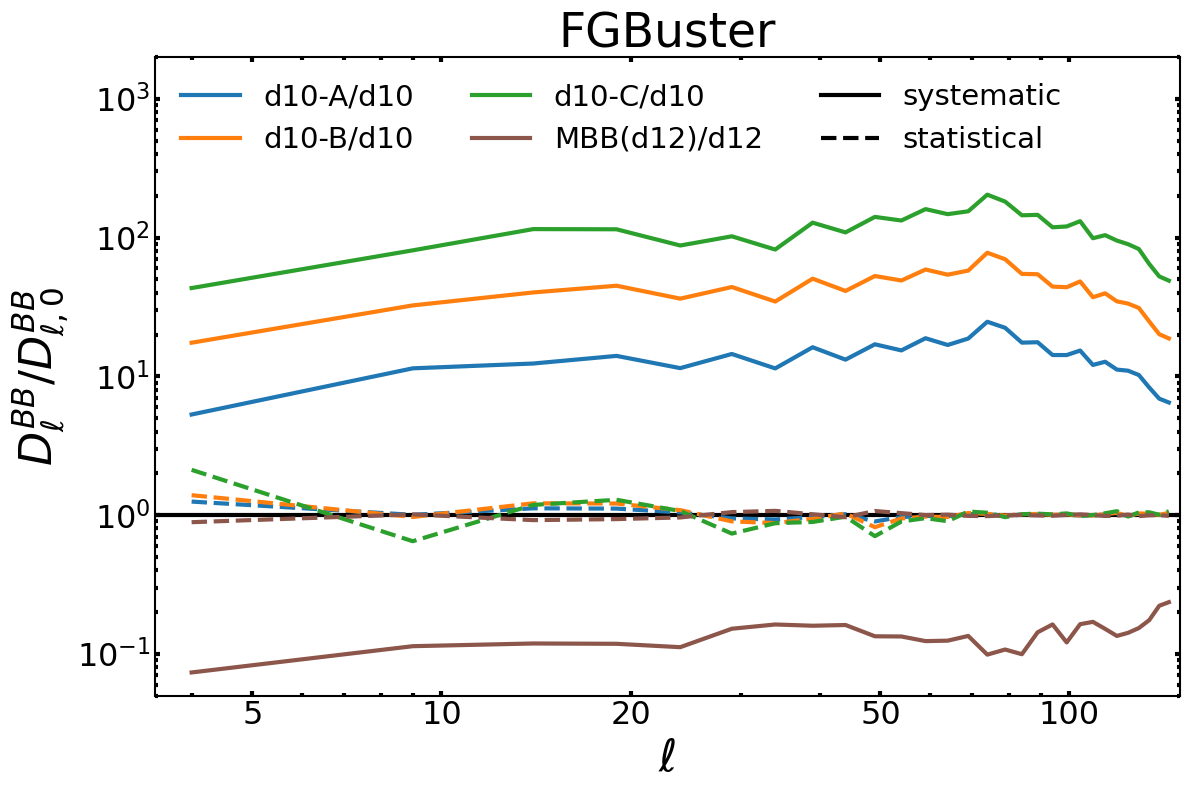}
    \includegraphics[width=0.97\columnwidth]{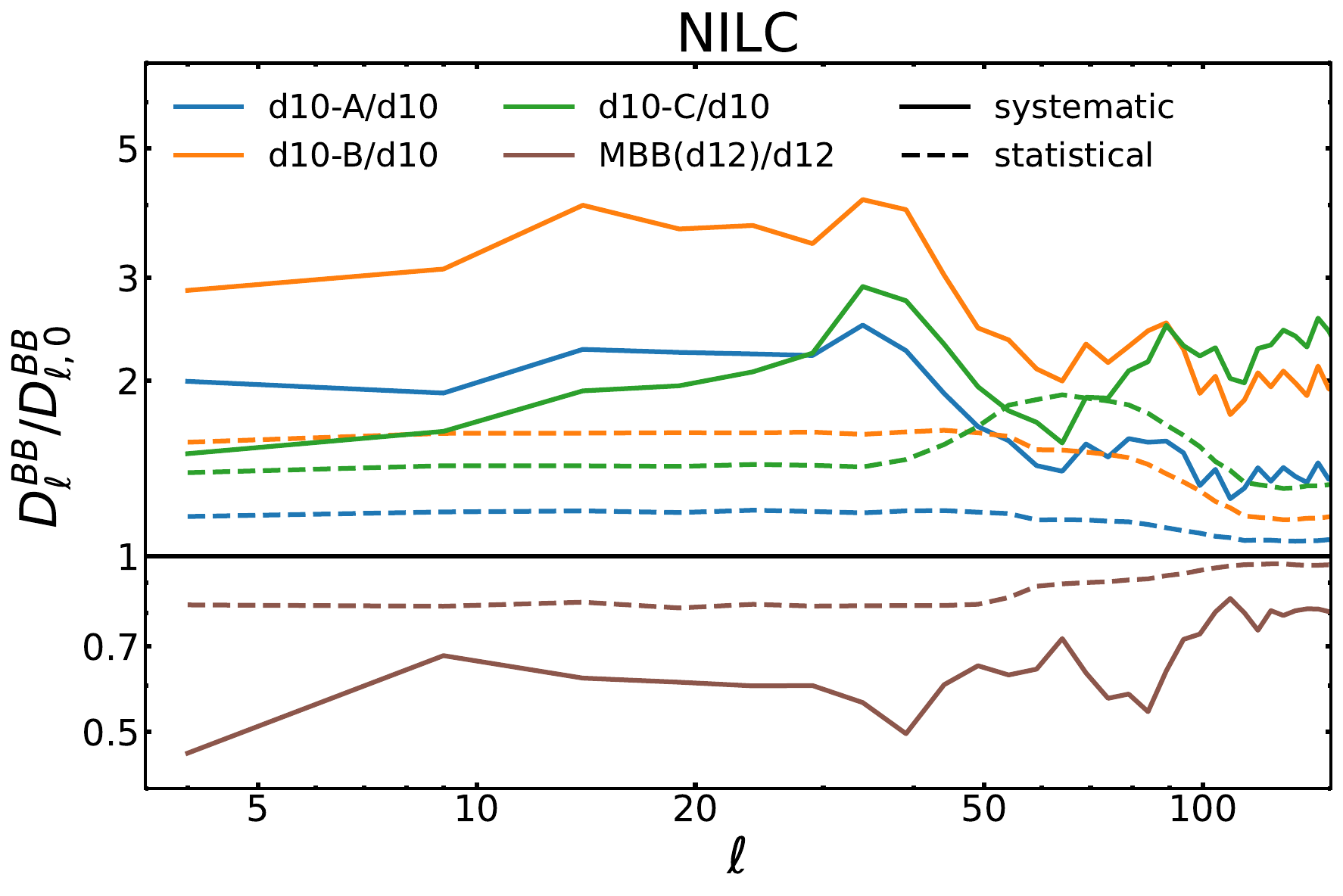}
    \caption{ 
    The ratio $\mathcal{D}_\ell^{BB}/\mathcal{D}_{\ell,0}^{BB}$ compares the angular power spectra of the $B$-mode systematic (solid lines) and statistical (dashed lines) residuals after component separation in the case of a model with intra-pixel complexity (\dten{}-A, \dten{}-B, \dten{}-C and \dtwelve{}) with those for their counterpart without intra-pixel complexity (\dten{} and MBB(\dtwelve{})). Results are shown for the parametric FGBuster and blind NILC pipelines in the left and right panel, respectively.
    All angular power spectra are computed excluding the Galactic plane with the \planck{} HFI \texttt{GAL70} mask which retains $70\%$ of the sky.
    }
    \label{fig:comp-sep}
\end{figure*}

The outcome of component separation is assessed through the computation of angular power spectra of systematic and statistical residuals, with the former biasing the final estimate of the CMB power spectrum, while the latter contributing to its overall uncertainty\footnote{The residuals are obtained differently for the two pipelines: in NILC, by combining the input foreground (systematic) and noise (statistical) multifrequency maps with the weights; in FGBuster, systematic residuals are derived by running the pipeline on foreground-only simulations, while statistical ones by subtracting the systematic to the total residuals.}. To avoid the most contaminated regions along the Galactic plane, the \planck{} HFI \texttt{GAL70}, which retains $70\%$ of the sky, is adopted. At this stage, we focus only on the analysis of the $B$-mode power spectrum. 
Since our goal is to quantify the impact of the intra-pixel complexity on the component-separation methodologies, in Fig.~\ref{fig:comp-sep} we display the spectral amplitude of the residuals for each model featuring 3D complexity -- models A, B and C (noted respectively \dten{}-A, \dten{}-B, \dten{}-C) and \texttt{d12} -- relatively to that obtained in the case of the analogous model but containing solely inter-pixel complexity -- \texttt{d10} and ${\rm MBB}(\dtwelve{})$. We use the suggestive notation ${\rm MBB}(\dtwelve{})$ to refer to overall MBB factor ($\alpha=0$) inferred from the \dtwelve{} model in Sect.~\ref{sec:d12} (also noted $\barP$). For completeness, the angular power spectra of the $B$-mode residuals after component separation are shown for both FGBuster and NILC for all the different models in Fig.~\ref{fig:comp-sep-full}. 

Quite interestingly, the two component separation methods display very different results. The residuals for the parametric method FGBuster follow exactly the hierarchy predicted in the previous section. Indeed, models A, B, and C feature increasing amplitude of residuals accordingly to the trend of the proxies of dust complexity $\Delta_\ell^{BB}$ and $r_\nu^{E/B}$ displayed in Fig.~\ref{fig:Delta-reB-d12}. Specifically, Model C, which maximises these proxies, has the largest systematic residuals at all scales, on average around 60 times greater than \dten{}'s at the spectrum level. 
We further observe that \dtwelve{} residuals are around one order of magnitude greater than the ones of ${\rm MBB}(\dtwelve{})$, implying that the additional intra-pixel complexity of \dtwelve{} has a significant impact on the parametric component separation, which could not be fully anticipated by looking solely at the proxies of complexity (Fig.~\ref{fig:Delta-reB-d12}).

The NILC presents a different -- if not opposite -- story. Model C, expected to be the most complex according to the proxies, does not lead anymore to the highest residuals -- both statistical and systematic -- on the largest scales.
Model B is by far the most complex, with around three times greater residuals than \dten{}. We further note that, here again, a significant difference exists between \dtwelve{} and ${\rm MBB}(\dtwelve{})$.

We can attempt at interpreting these results as follows. \reviewcorrect{The hierarchy of the residuals for parametric methods can be rather straightforwardly deduced from the classical proxies of foreground complexity as parametric methods are directly sensitive to the effects of mixing which introduce deviations from the canonical SED assumed in the parametric model.}{First, it seems that the hierarchy of the residuals for parametric methods follows the hierarchy of the classical proxies of foreground complexity (the decorrelation and the $E/B$ ratio). This fact can be explained as parametric methods are directly sensitive to deviations from the canonical SED which they assume to model the foreground signal. Such deviations are a direct consequence of the mixing (through both the SED distortions and the spectral rotation of the polarisation angle) and are probed directly by the proxies of complexity.} As discussed in Sect.~\ref{sec:d12}, the imaginary part of the moments, despite having small amplitudes, can generate a pernicious complexity by entangling $Q$ and $U$ as well as $E$ and $B$, explaining why the model C, with the largest imaginary moments, has both the largest proxies of complexity and the highest residuals for the parametric approach. 

On the other hand, in regard of Fig.~\ref{fig:comp-sep}, we should conclude that minimum variance methods are not overall strongly sensitive to \review{the effects probed by} the proxies of complexities. This can be justified as the application of such methods only requires the computation of the input multifrequency covariance. Therefore, any significant foreground distortion term contributing to the covariance structure is properly addressed. Similarly, some $E-B$ leakage and rotation of the polarisation angle is expected to have a smaller impact as these methods perform their analysis only on $B$-mode maps.
However, we find instead that the NILC residuals correlate with the 'relative-power' $\mathcal{C}_\ell^{BB}(100)/\mathcal{C}_\ell^{BB}(353)$, displayed in Fig.~\ref{fig:relative-power}. By definition, this observable quantifies how much power is added by the moments at the frequency $\nu=100\,$GHz, away from the reference frequency $\nu_0=353\,$GHz where the moment contribution is null. More precisely, the relative power is driven by the modulus of the moment maps, whose main contribution comes from their real parts, which are significantly larger in amplitude than their imaginary parts (see Appendix ~\ref{app:mom-maps}). Therefore, model B, for which the moments are rigidly amplified by a constant factor, turns out to have the largest moment modulus and thus the largest relative power, and it turns out to be the most complex for NILC. We can then speculate that relative power provides a reliable proxy to assess the complexity of models regarding to NILC. We note that if such a claim revealed to be correct, it would explain a posteriori why the relative power provides such a powerful proxy in order to proceed to build clusters in advanced versions of NILC as MCNILC  \citep{Carones2022b}. 

We note that the excursion of the residuals when some 3D complexity is injected is much higher for parametric methods ($\mathcal{O}(10)$) compared to the ones of minimum variance ($\mathcal{O}(1)$). This difference can be tentatively explained by the fact that parametric methods perform very efficiently when the parametric model is close to the data signal (d10, MBB(d12)) while their performance is greatly degraded when the signal deviates strongly from the assumed SED model. 3D complexity  thus represent a major challenge for these methods as it introduces such deviations in each pixel (as in the cases of \dtwelve{}, \dten{}-A, \dten{}-B and \dten{}-C). On the other hand, minimum variance approaches do not assume a model, and therefore, their performance does not depend on the origin of the complexity, that is, whether the complexity arises from 2D (inter-pixel) or 3D (intra-pixel) distortions.
Nevertheless, all the previously discussed results aim only at investigating the hierarchy of the impact of the different models after application of component separation. We are not assessing here the general capability of component separation to deal with the most complex models. A more careful application of FGBuster as the one proposed in \cite{FGCluster}, alternative parametric methods based on moments \citep{Ichiki2019,Azzoni2020,Vacher2022a} or minimally informed \citep{Leloup2023,Morshed2024}, as well as extended versions of NILC as MC-NILC or oc-MILC \citep{RemazeillesmomentsILC, Carones2022,Carones2024} might be able to provide an effective reconstruction of the CMB signal for such challenging models including intra-pixel complexity. For the same reason, we do not take some realistic effects into account in the simulations, such as the scanning strategy or instrumental systematics, despite the latter have been shown to be deeply coupled with the background foreground model \citep{NPIPE, sroll2, cosmoglobe}. Still, we conclude that special care has to be given on attempts at assessing the complexity of a foreground model since such a question is deeply dependent on the component separation method under consideration. Different methods are sensitive to different facets of foreground complexity, and the moment coefficients provide a powerful tool to separate and study these different aspects. Still, we were able to propose novel alternative models -- such as model B -- which present a significant challenge for both parametric and minimum variance methods while remaining compatible with data.

\review{We also note that the present analysis assumed simple Gaussian white noise without any additional systematic effect. A more realistic analysis would also have to deal with the complex coupling between systematic effects (calibration of the polarisation angle, far side lobe leakages, gain uncertainties ...)  and the additional complexity added to the foreground. In many cases, the foreground and the systematic complexity become entangled, such that it becomes an extremely difficult task to separate them.}

\section{Discussion and conclusion}
\label{sec:conclusion}

In this work, we have demonstrated that the use of moment maps provides a powerful tool to quantify and model the complexity of the foreground SED models in a minimal and simple way. In principle, the complexity of virtually any foreground model currently used by the CMB community can be re-expressed by a set of moment maps. This complexity can then be amplified or reduced in a consistent way by changing the amplitude or patterns of these maps as desired. This approach allows one to perform careful studies of the link between the complexity contained in the maps and the impact of this complexity on the power spectra and component separation. In our analysis, we provided a first application of this concept by accurately reproducing the most complex \texttt{PySM} dust model (\dtwelve{}) in terms of moment maps. We were able to reproduce such a model at the one-per-ten-thousand level using five moment maps at the second order of the expansion. In doing so, we learned that despite their small amplitudes, the imaginary parts of the first order moments are the main drivers of $B$-mode decorrelation. The expansion, however, must be pushed up to the second order to accurately recover the dependence of the $E/B$ ratio with frequency. In a second step, we re-used \dtwelve{}'s moment maps to inject some additional complexity into each pixel of a simpler foreground model ($\dten{}$), simulating the presence of mixing along the line of sight. This injection was done in a self-consistent fashion, which can be interpreted in terms of ISM physics. By scaling these moment maps, we were able to build extremely complex models with regard to the proxies of complexity given by decorrelation of the $B$-modes $\Delta_\ell^{BB}$ and spectral dependence of the $E$-to-$B$ ratio $r^{E/B}_\nu$.
As such, we proved that all the models currently used by the CMB community (including \dtwelve{}) contain much milder distortions coming from the third dimension than what can be allowed by data. It is thus crucial to assess whether component separation methods are robust enough to handle these new complex scenarios. 

Using component separation on the newly constructed models, we saw that the additional complexity could have a significant -- but not insurmountable -- impact on the quality of the reconstruction of CMB polarisation fields. Quite interestingly, different component separation methods are not sensitive to the same kind of complexity in the same way to the point that they do not agree on what is the most complex model. Parametric methods, such as FGBuster, are very sensitive to deviations of the signal from a canonical SED. They are thus more impacted by models having significant decorrelation and frequency dependence of the $E/B$ ratio, which can be largely driven by the imaginary part of the first order moment maps, as discussed in Sect.~\ref{sec:d12}. On the other hand, minimum variance methods are pretty much insensitive to this complexity but are more sensitive to the total power added by the moment maps to the foreground signal. They are thus sensitive to the modulus of the moment maps regardless of whether the complexity is coming from their real or imaginary part.
A lesson therefore learned from this investigation is that caution is required when assessing the `complexity' of a model, as the different quantifiers of this complexity can depict contradictory stories. 

The models studied here must be understood as toy models that allow for the investigation of the relevance of the moment expansion formalism for foreground modelling as well as the impact of some possible additional complexity in the third dimension. However, it is clear that none of these models can pretend to be physical and representative of the actual or expected complexity of the sky. Clearly, the moment maps used in this work do not contain the non-Gaussian complex structure characteristic of the cosmic dust signal. While it has been recently suggested by \cite{Abril-Cabezas2024} that component separation for the CMB is largely insensitive to such non-Gaussianities, the impact of non-Gaussian moments has never been discussed as of yet. Powerful tools such as Minkowski functionals \citep{Carones2022,2024JCAP...01..039C} and more generically scattering transforms \citep{Allys2019,Delouis2022,Mousset2024} could be used to study and model these non-Gaussianities on the sphere. 
\review{Furthermore, we established the validity of our models solely based on the differences of their polarised angular power-spectra power with the measurement of the \planck{} satellite. Doing so ensured that the sum of the inter-pixel complexity and the added intra-pixel complexity is statistically compatible with current data. However, a refined-map based analysis could certainly reject the existence of certain deterministic patterns injected in the models. Such an analysis, however, would be difficult to perform consistently, as the current CMB data is not sensitive enough to allow one to distinguish the inter- and the intra-pixel complexities. As such, our analysis can only claim to explore the upper limits on the total complexity of the foregrounds and its impact, and it does not build any new realistic and fine tuned models. The addition of other data than the CMB observations would be necessary to build realistic moment maps and finely distinguish the plane of sky and the line of sight contributions.} In the long term, measurements such as the ones of \cite{Zelko2022} and \cite{Pelgrims2024} should be used to place refined bounds and build more realistic and reliable moment maps representing the actual complexity contained in the third dimension. 
Until then, one remains free to fine-tune as desired the morphologies and amplitudes of the added moment maps in order to introduce and study the effect of non-Gaussianities or any other kind of complexity.

The present analysis focused on the impact of thermal dust on the recovery of the primordial $B$ modes. However, it would be straightforward and beneficial to apply such a methodology in other contexts. Firstly, the complexity of any type of foreground signal other than dust can be similarly interpreted, decomposed, and amplified using moment maps. A natural further step would then be to repeat the present work focusing on the synchrotron emission. 
Secondly, additional line-of-sight complexity must also impact at some level the recovery of the CMB $EB$ spectra. To assess this impact, one must consider the addition of the moment spectral complexity to a model containing a significant $EB$ foreground signal, such as the one proposed in \cite{Hervias-Caimapo2024}, and see whether or not analysis similar to the ones performed in \cite{Diego-Palazuelos2022b,Jost2023} remain robust in such maximally complex scenarios.
Thus, more general studies providing a systematic link between different sets of moment maps and the resulting 3D complexity for various types of foregrounds and analysis are much desired and will be provided in future works. 

\begin{acknowledgements}

LV would like to thank Carlo Baccigalupi for constant support, comments and useful discussions through the entirety of this project.
LV and AC acknowledge partial support by the Italian Space Agency LiteBIRD Project (ASI Grants No. 2020-9-HH.0 and 2016-24-H.1-2018), as well as the RadioForegroundsPlus Project HORIZON-CL4-2023-SPACE-01, GA 101135036. 
MR acknowledges the support of the Spanish Ministry of Science and Innovation through projects~PID2022-139223OB-C21 and PID2022-140670NA-I00, funded by the Spanish MCIN/AEI/10.13039/501100011033/FEDER, UE.

\end{acknowledgements}
\bibliographystyle{aa}
\bibliography{bi}
\begin{appendix}
\section{A word on normalisation \label{app:normalisation}}

The moments coefficients can be defined using different choices of normalisations. Such a choice must be made carefully depending on the problem at stake, as divergences can appear in the case of the polarised signal. Writing explicitly the modified blackbody function, the expansion considered in this work (given by Eq.~\eqref{eq:mom-exp-MBB}) becomes
\begin{align}
    \langle\mathcal{P}_\nu \rangle = \overline{\mathcal{A}}\left(\frac{\nu}{\nu_0}\right)^{\overline{\beta}}B_\nu(\overline{\tempd})\times\Bigg( 1 + &\mathcal{W}_1^{\betad}\ln\left(\frac{\nu}{\nu_0}\right) +\mathcal{W}_1^{\tempd} \Delta\Theta_\nu\nonumber+\cdots\Bigg),
\end{align}
where in the discrete case, the auto spin moments are defined as $\Walphap= \sum_i \mathcal{A}_l(p_l-\bar{p})^\alpha/\mathcal{A}$ with $p\in \{\beta,\tempd\}$ and the sum is ranging on the number of layers.

Alternatively, one can rewrite this expansion as
\begin{align}
    \langle\mathcal{P}_\nu \rangle = \left(\frac{\nu}{\nu_0}\right)^{\overline{\beta}}B_\nu(\overline{\tempd})\times\Bigg( \overline{\mathcal{A}} + &\widetilde{\mathcal{W}}_1^{\betad}\ln\left(\frac{\nu}{\nu_0}\right) +\widetilde{\mathcal{W}}_1^{\tempd} \Delta\Theta_\nu\nonumber+\cdots\Bigg),
\end{align}
renormalising the moments as $\widetilde{\mathcal{W}}_\alpha^p= \sum_i \mathcal{A}_l(p_l-\bar{p})^\alpha$. This second choice of normalisation should in general be preferred as the first one can diverge in the regions where $\mathcal{A}\to 0$. Discussions on this point can be found in \cite{Vacher2022b} and \cite{Vacher2023}.

In the present work, we preferred the first choice of normalisation, as it allows the moment maps to 'forget' about the structure of the original map contained in $\mathcal{A}$, such that they can be applied identically to any other maps. While this is clear when comparing the definitions of $\mathcal{W}^p_\alpha$ with $\widetilde{\mathcal{W}}^p_\alpha$, we justify further this point in Appendix~\ref{app:universality}. Furthermore, using the first choice of normalisation, moments are unitless coefficients associated with clear interpretations which are lacking using the second choice. We discuss how to provide such interpretations for the case of \dtwelve{} in Appendix~\ref{app:mom-maps}.

However, our choice of normalisation can reveal problematic due to possible divergences in regions where the polarised signal is close to zero. As a sanity check, we verified that no outliers points were caused by such divergences and we reproduced our application to the \dtwelve{} model presented in Sect.~\ref{sec:d12} using the second choice of normalisation, leading to identical results.

Nevertheless, we note that divergences cannot be evaded when computing the pivot, which expression does not depend on the choice of normalisation (Eq.~\eqref{eq:p_bar}). In this work, we avoided such a problem by forcing the pivot to take the value of intensity when its value was outside of a given interval (see footnote \ref{foot:pivot}).

\section{Universality of the moment maps \label{app:universality}}

\begin{figure*}[h!]
    \centering   
    \includegraphics[width=\textwidth]{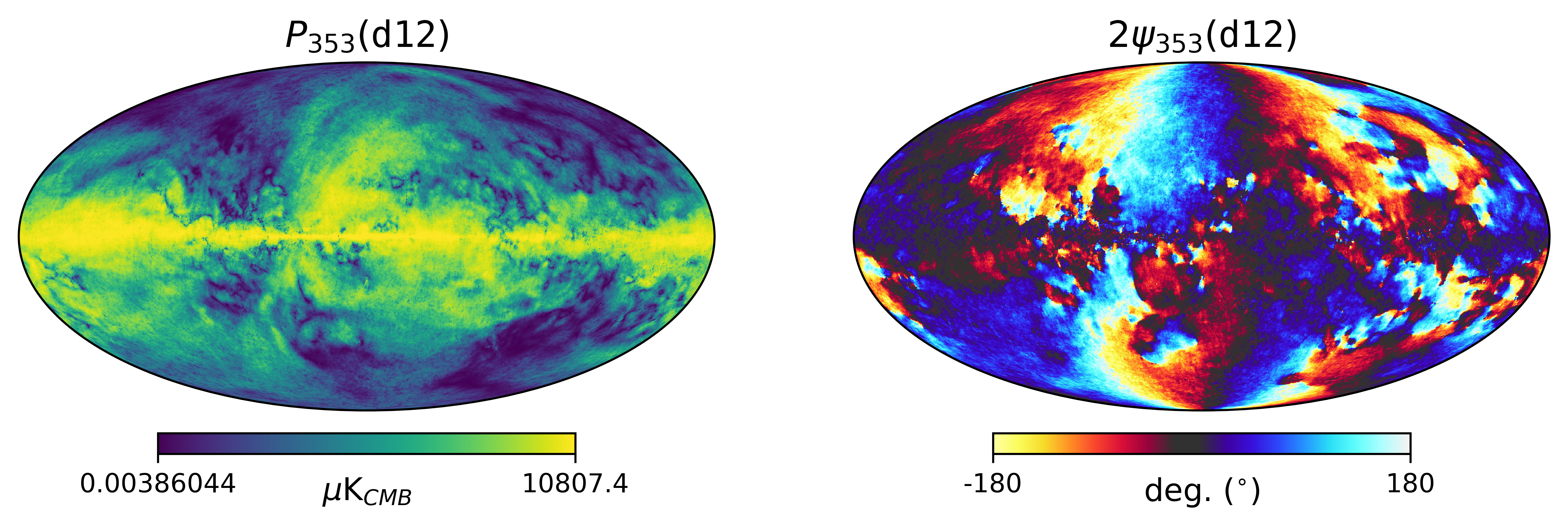}
    \includegraphics[width=\textwidth]{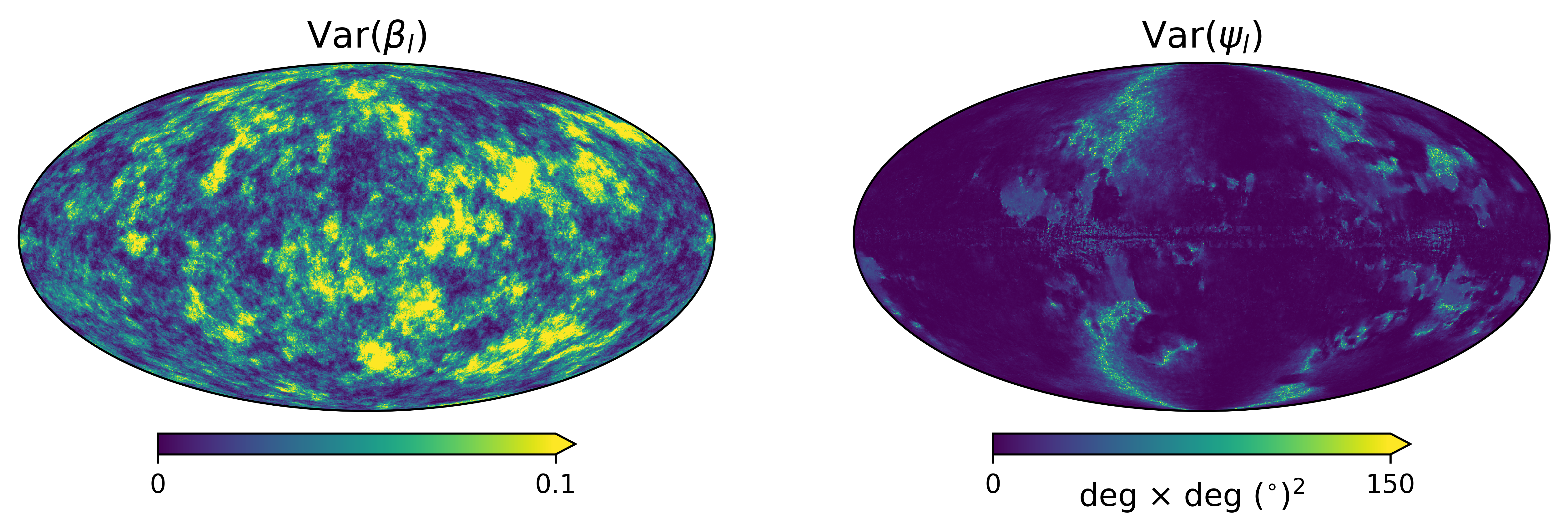}
    \caption{\review{Upper panels: Polarised intensity (upper left) and absolute value of the polarisation angle (upper right) of the {\tt d12} model observed at $\nu_0=353$ GHz. Lower panels: Variances of the spectral index $\beta_l$ (lower left) and (circular) variance of the polarisation angles $\psi_l$ contained in each pixel of the {\tt d12} at the native resolution. }}
    \label{fig:var2D-3D-d12}
\end{figure*}

We computed our moments from the \dtwelve{} maps using Eq.~\eqref{eq:moments}
and the pivots computed with Eq.~\eqref{pbar-MBB}. When applying these moments directly to the other model \dten{}, one can worry that doing so will carry along some inter-pixel complexity of the original map of $\dtwelve{}$ and not only information about the complexity contained within the line of sights.

To do things properly, we tried to find some other set of parameters $\tilde{\mathcal{A}}_l$ and $\tilde{p}_l$ satisfying the properties 
\begin{align}
&\sum_l \tilde{\mathcal{A}}_l = \overline{\mathcal{A}}(\dten{})\label{eq:B.1}\\
&{\rm Re}\left(\frac{\sum_l\tilde{\mathcal{A}}_l\tilde{p}_l}{\sum_l \tilde{\mathcal{A}}_k}\right)= \bar{p}(\dten{})\label{eq:B.2}
\end{align} 
and which could model the 3D complexity for \dten{} with the statistics of the layers from \dtwelve{}. Equation~\eqref{eq:B.1} asks that the sum of all the new amplitudes $\tilde{\mathcal{A}}_l$ should gives back the total amplitude of \dten{} instead of of \dtwelve{}, while the second asks that the pivot of the moment expansion should be the spectral parameters of \dten{} and not the ones of \dtwelve{}.

First, we can redefine the amplitudes in the layers as $\tilde{\mathcal{A}}_l = \overline{\mathcal{A}}(\dten{})\mathcal{A}_l/\overline{\mathcal{A}}(\dtwelve{})$ which satisfies Eq.~\eqref{eq:B.1}. Due to our choice of normalisation, it is clear that such a constant rescaling will have no impact on the value of the moments nor the pivots. 

Then, we can redefine the parameters $\tilde{p}_l = p_l + \bar{p}(\dten{}) - \bar{p}$ which satisfies Eq.~\eqref{eq:B.2}. Since $\tilde{p}_l$ are different from $p_l$ only by a constant shift, all their statistical properties but the mean should remain unchanged, thus preserving the original intra-pixel complexity.

The values of the moments with this rescaling will then be
\begin{align}
\Walphap&=\frac{\sum_l\tilde{\mathcal{A}}_l(\tilde{p}_l-\bar{p}(\dten{}))^\alpha}{\sum_l \tilde{\mathcal{A}}_k}\\
&=\frac{\sum_l\mathcal{A}_l(p_l + \bar{p}(\dten{}) - \bar{p}-\bar{p}(\dten{}))^\alpha}{\sum_l \mathcal{A}_k}\\
&= \frac{\sum_l\mathcal{A}_l(p_l - \bar{p})^\alpha}{\sum_l \mathcal{A}_k},
\end{align}
which is nothing less than the value of the original moments computed from \dtwelve{}. As such, a given set of moment maps can be associated to any sky map regardless of its amplitude and spectral parameters as the moment maps contains only information about the third dimension. \review{This last point can be verified by comparing the moment maps such as the one of Fig.~\ref{fig:W2b-d12} with some of the inter-pixel and intra pixel properties of {\tt d12} displayed in Fig.~\ref{fig:var2D-3D-d12}. The moment maps thus normalised do not correlate with the polarised amplitude or polarisation angles of the \dtwelve{} maps, but only quantify the 3D complexity contained in each pixel.}

\section{Interpreting \dtwelve{}'s moment maps \label{app:mom-maps}}

All the moment maps appearing as $\Walphap$ in Eq.~\eqref{eq:mom-exp-MBB} are spin-2 fields on the sphere, or simply put maps of complex numbers. They quantify the complexity arising from the co-variations of the spectral parameters and the angles along the line of sight. The total map in each pixel is thus the product of $\barP$ which is a perfect MBB map with varying $\overline{\betad}$ and $\overline{\tempd}$ in each pixel and the correction term added by the moments arising from additional complexity in the line of sight (or in 2D inside the pixel). Each moment of order $\alpha$ of the parameter $p$ is related to the statistical moment of order $\alpha$ (mean, \reviewcorrect{standard deviation}{variance}, skewness...) of the distribution of $p$ along the line of sight.
\begin{figure*}[h!]
    \centering   
    \includegraphics[width=\textwidth]{modphase-mom-d12.png}
    \caption{Full sky maps of the \reviewcorrect{real and imaginary parts of all the non-zero}{modulus and phase of all the} spin moments involved in the expansion of the $\dtwelve{}$ map at the second order. \review{We note that the phase of first order moments, not represented here, are strictly equal to $90^{\circ}$ on the whole sky, translating that they are pure imaginary numbers. They thus contribute maximally to the spectral rotation of the polarisation angles.}}
    \label{fig:allmom-d12-modphase}
\end{figure*}
\begin{figure*}[h!]
    \centering   
    \includegraphics[width=\textwidth]{allmom-d12-update.png}
    \caption{Full sky maps of the real and imaginary parts of all the non-zero spin moments involved in the expansion of the $\dtwelve{}$ map at the second order.}
    \label{fig:allmom-d12}
\end{figure*}
\begin{figure*}[h!]
    \centering   
    \includegraphics[width=\textwidth]{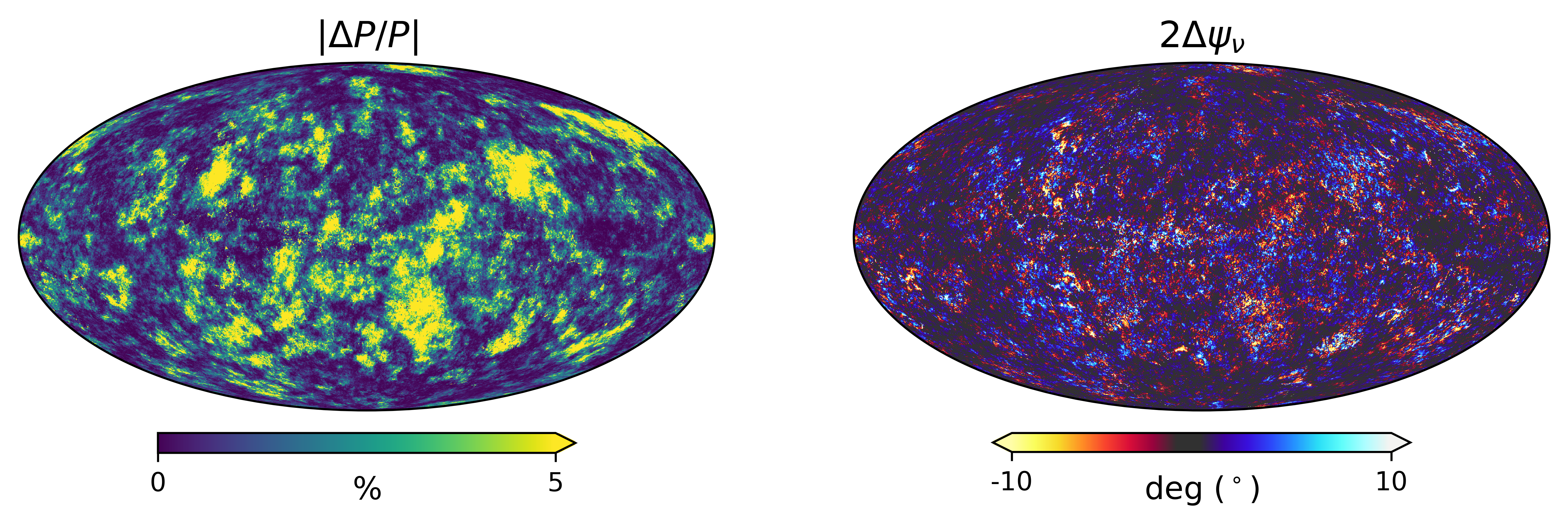}
    \includegraphics[width=\columnwidth]{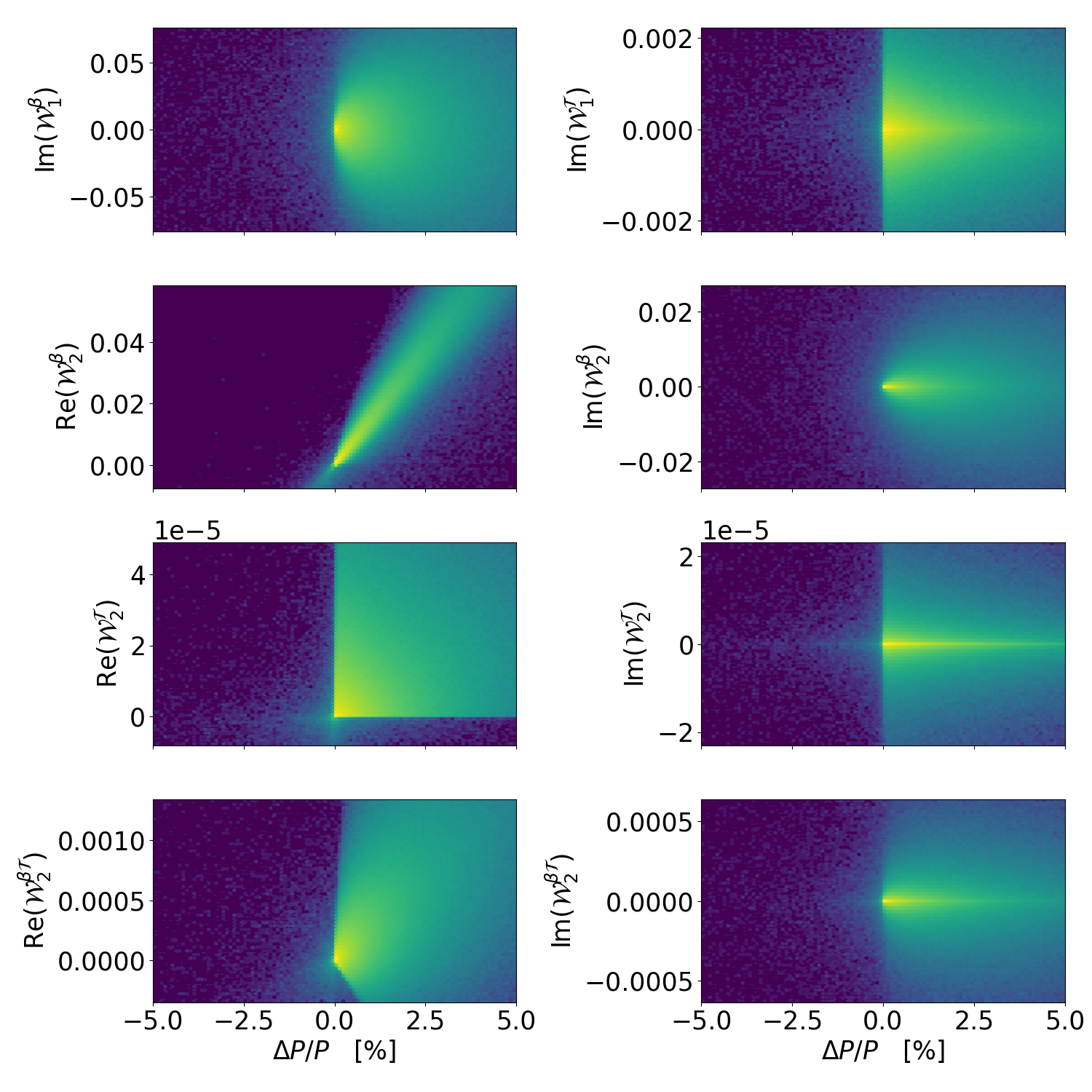}
    \includegraphics[width=\columnwidth]{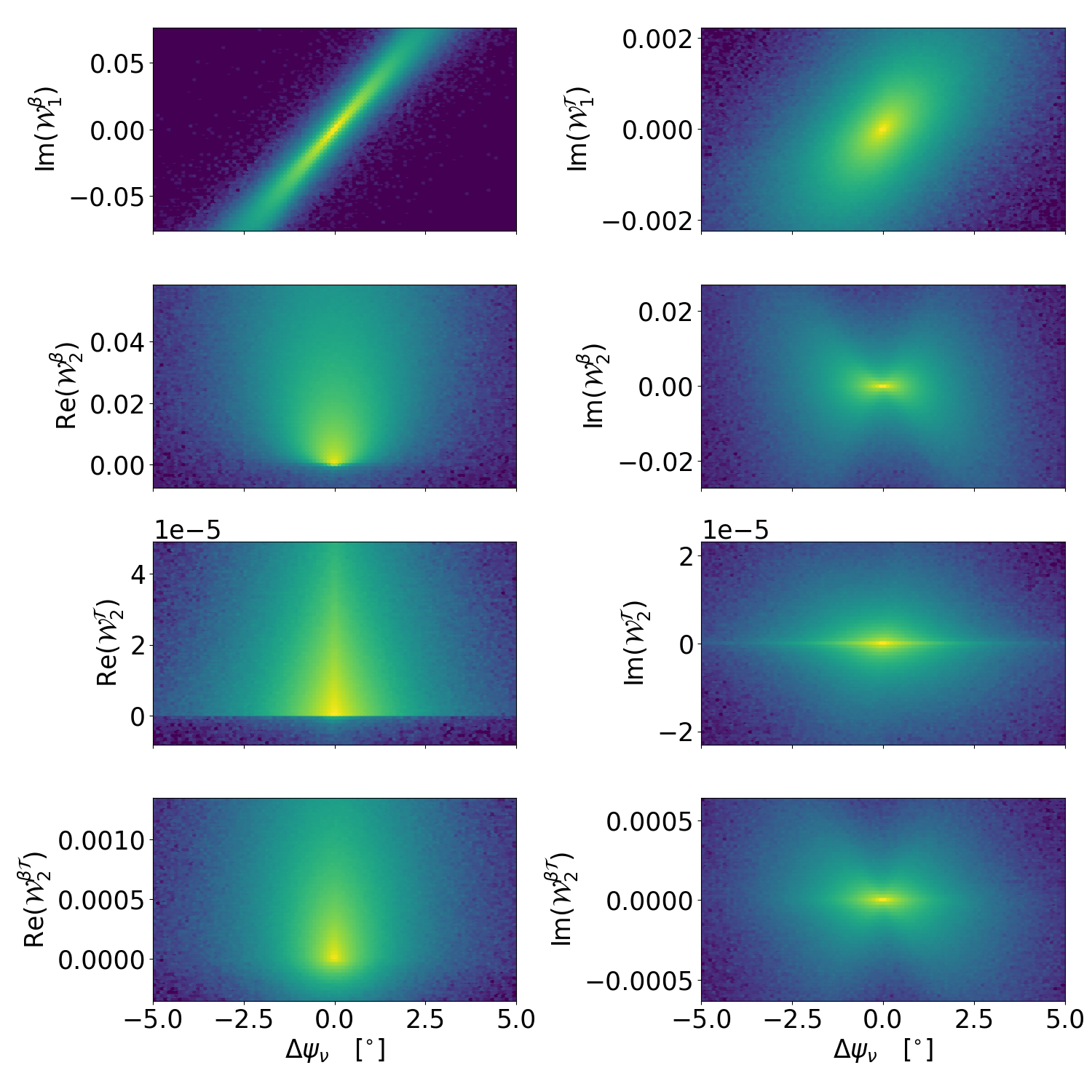}
    \caption{\review{Upper panels: Maps of the distortion of the polarised intensity $\Delta P/P$ (left) and the spectral rotation of the angles $\Delta \psi_\nu$ (right) for the \dtwelve{} model. Bottom panels:} Relation between the moments value in every pixel and the \reviewcorrect{spectral rotation of the angles $\Delta \psi_\nu$ (left) and the distortion of the polarised intensity $\Delta P/P$ (right) for the \dtwelve{} model}{the two quantities presented in the upper panels}. $\Delta \psi_\nu$ is defined as the difference of the polarisation angle in each pixel between the \dtwelve{} emission at $100$ and at $353$ GHz. $\Delta P/P$ is defined as $(P_{\tt d12}-\bar{P})/P_{\tt d12}$, where $P_{\tt d12}$ is the polarised intensity ($P=\sqrt{Q^2+U^2}$) of the original \dtwelve{} model, while $\bar{P}$ the one of the pivot modified blackbody of the expansion. }
    \label{fig:momd12-correl}
\end{figure*}
\begin{figure*}[h!]
    \centering   \includegraphics[width=\columnwidth]{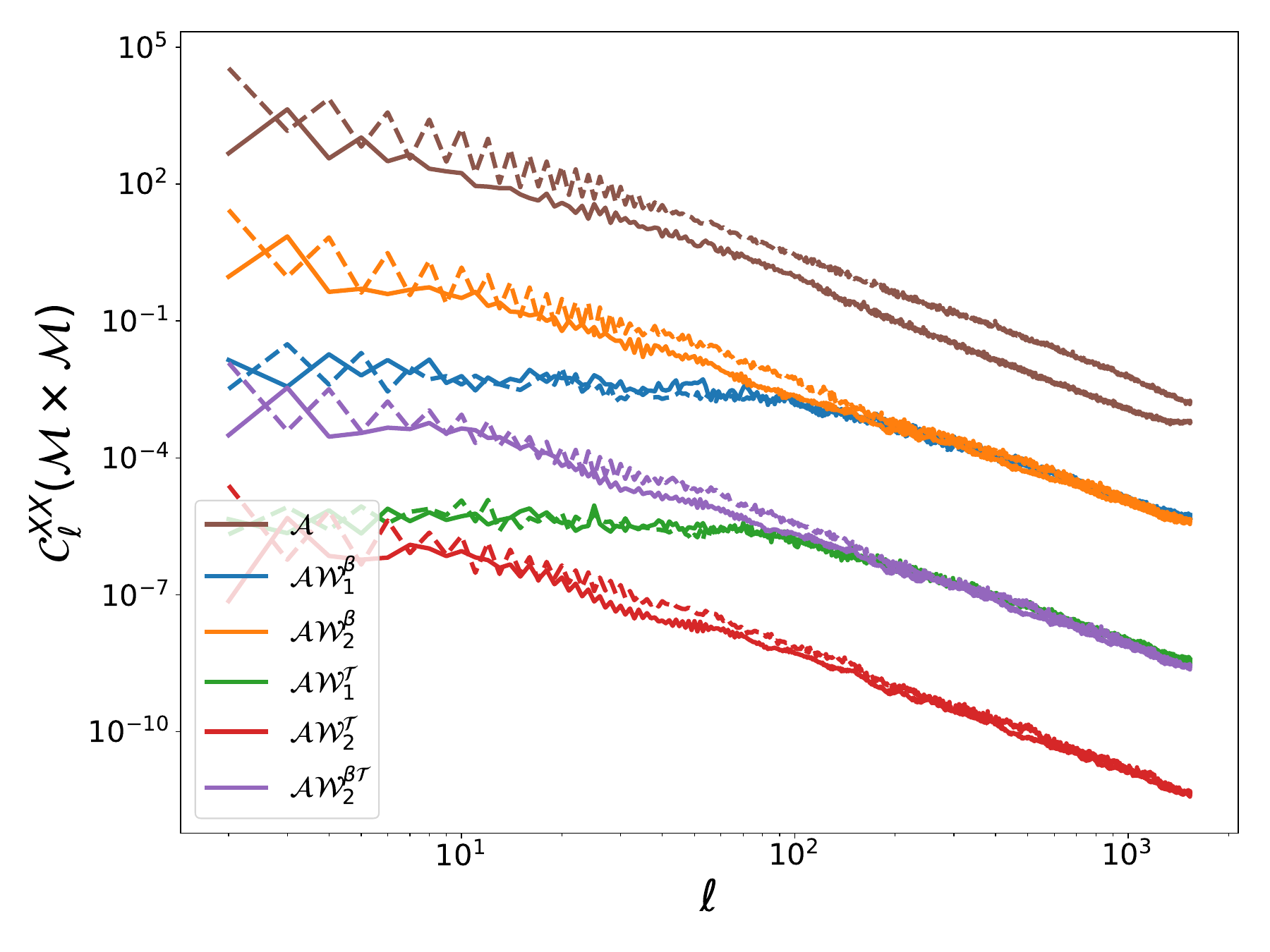}\includegraphics[width=\columnwidth]{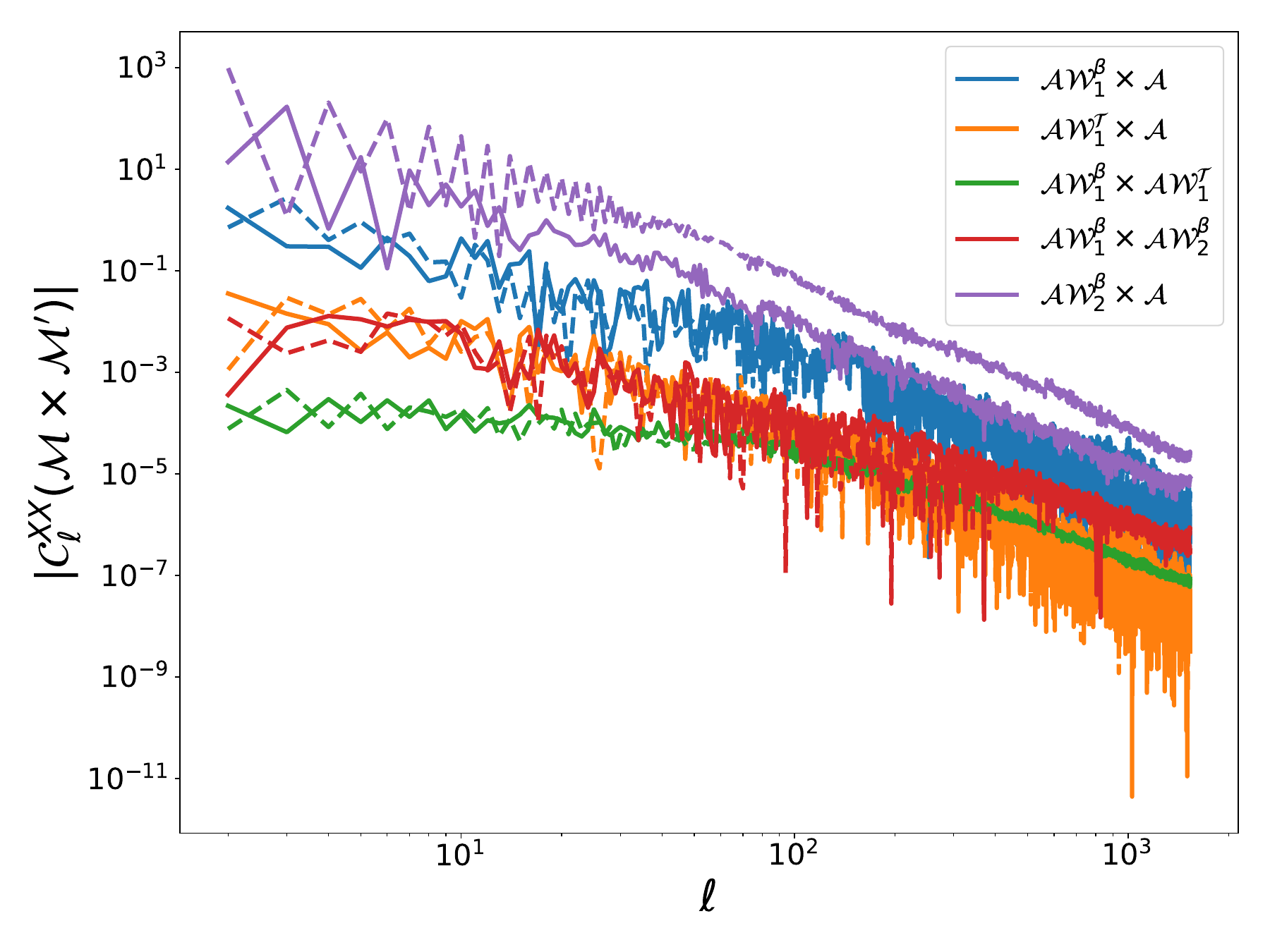}
    \caption{Full sky  angular power spectra of the moment maps of \dtwelve{}. All the auto-spectra of the expansion up to second order are represented on the left panel while the dominant terms of the (absolute values of the) cross-spectra on the right panel. The full lines represent the $BB$ spectra while the dashed lines represent the $EE$ spectra. \review{The spectra are expressed in units of $(\mu {\rm K}_{CMB})^2\cdot {\rm K}^{-2\alpha_{\mathcal{T}}}$ where $\alpha_{\mathcal{T}}$ is the order of the moment with respect to $\mathcal{T}$.}}
    \label{fig:cl-mom-d12}
\end{figure*}

The full sky maps of the \reviewcorrect{real and imaginary parts}{modulus and phase} of the spin moment maps present in the expansion of \dtwelve{} at second order can be found in \reviewcorrect{Fig.~C.1}{Fig.~\ref{fig:allmom-d12-modphase}, and their real and imaginary parts are displayed in Fig.~\ref{fig:allmom-d12}}. \reviewcorrect{All maps}{The maps of the real and imaginary parts (as well as the modulus)} present a Gaussian structure lacking the characteristic non-Gaussian features of the ISM. Strong similarities can be seen between the real parts \review{and modulus} of $\mathcal{W}_2^\beta$ and $\mathcal{W}_2^\tempd$, translating a strong correlation between $\beta$ and $\tempd$ also visible in the non-zero value of ${\rm Re}(\mathcal{W}_2^{\beta \tempd})$. The real moment maps display some 'spots' of negative regions which might appear as pathological. These spots are associated with regions of high variations of $\psi$ along the line of sight in which the moments can rotate and acquire a significant phase and randomly acquire a negative real component (as an Euclidian vector in a Cartesian frame pointing towards the $x$ direction $\vec{v}=v_x\vec{u}_x$ acquires general components $v_x'$ and $v_y'$ after rotation, which can be either positive or negative). \review{These negative regions are regions where the phase of the moment maps has a high variability (Fig.~\ref{fig:allmom-d12-modphase}) and are strongly correlated with regions of large variance of $\psi$ along the line of sight (bottom panel of Fig.~\ref{fig:var2D-3D-d12}). }

In the case where no variation of the polarisation angles exist along the line-of-sight, the moments are purely real maps. In that case the modified blackbody map $\barP=\overline{Q}_\nu+ \i \overline{U}_\nu$ is only multiplied by real functions of the frequency, modifying the SED in each pixel without any transfer of $Q$ into $U$.

On the other hand, the imaginary parts of the moments result from the combined variation of large deviations of $p_l$ from $\bar{p}$ and variation of the different polarisation angle along the line of sight $\psi_l={\rm Arg}(\mathcal{A}_l)/2$. Non-zero imaginary moments will unavoidably result in a transfer of $\overline{Q}$ into $\overline{U}$ inducing a greater variation of the polarisation angles with frequency for the total map along with a modification of the polarised SED.

As illustrated in Fig.~\ref{fig:momd12-correl}, the presence of the $\beta$ moments are the main driver of the complexity contained within \dtwelve{}. The coefficient ${\rm Re}(\mathcal{W}_2^\beta)$ presents the strongest correlation with the deviation of the polarised intensity from a modified blackbody in each pixel ($\Delta P/P$). As such, the large standard deviation of $\beta$ in each pixel is the main driver of the SED distortions. The real parts of the other moments, ${\rm Re}(\mathcal{W}_2^\tempd)$ and ${\rm Re}(\mathcal{W}_2^{\beta\tempd})$ also present a significant correlation with $\Delta P/P$ as well as ${\rm Im}(\mathcal{W}_2^\beta)$ and ${\rm Im}(\mathcal{W}_1^\beta)$ such that the other moment terms also contribute in a lesser extend to the value of $\Delta P/P$.

On the other hand, the values of 
${\rm Im}(\mathcal{W}_1^\beta)$ and ${\rm Im}(\mathcal{W}_1^\tempd)$ are strongly correlated with the variation of the total polarisation angle $\Delta \psi_\nu$ between 100 and 353 GHz. This is no surprise as these coefficients are expected to represent the leading order terms for the spectral rotation of $\psi$ as discussed in \cite{Vacher2022b,Vacher2023}. As expected from our previous discussion, the real part of the moment coefficients are not significantly correlated with $\Delta \psi_\nu$ while the imaginary parts ${\rm Im}(\mathcal{W}_2^\beta)$ and ${\rm Im}(\mathcal{W}_2^\tempd)$ display a `butterfly' shape that allows for possible correlation.

The connection between the moment maps and the proxies of complexity as $\Delta^{BB}_\ell$ and $r^{E/B}_\nu$ is not as straightforward in the general case.
The final impact will result from a combination of the original map and the moment angular power spectra.
We note that, in the normalisation used here, even constant moment maps would have a significant impact, as each term in Eq.~\eqref{eq:mom-exp-MBB} is multiplied by the original map $\barP$. This choice of normalisation for the moments is further justified and discussed in Appendix~\ref{app:normalisation}.

The $BB$ and $EE$ angular power spectra of the \dtwelve{} moment maps multiplied by $\mathcal{A}$ are displayed on Fig.~\ref{fig:cl-mom-d12}.  They all behave as inverse power laws in $\ell$, which is characteristic of foreground spectra. Here again, we see clearly that the first and second order terms in $\beta$ are dominant. The difference between the $E$- and the $B$-spectra are responsible for the values of $r^{E/B}_\nu$. We see that this difference is larger for second order terms than the first first orders, thus explaining the results on the right panel of Fig.~\ref{fig:Delta-reB-d12}. Just like lensing, moments will mix the $E$- and $B$-components of the signal, such that some of the dust $E$ modes will leak into the $B$ modes.

The exact link between the spin-moment maps and spectra and their impact on the total dust spectra was partially discussed in \cite{Vacher2023} and a more detailed analysis linking the moment maps to their impact on the proxies of foreground complexity in different special cases would be needed and left for future work. Such a refined analysis would be greatly beneficial for the interpretation of the results of parametric component separation method based on moments in harmonic space as \cite{Azzoni2020} and \cite{Vacher2022a}. 

\section{Derivation of \planck{} error bars \label{app:error-bars}}

The \planck{} \texttt{NPIPE} \citep{NPIPE} data release is accompanied by 600 high-fidelity Monte-Carlo simulations that include CMB, instrumental noise, foregrounds and systematics, both for full-mission and detector-set (A, B splits) maps. The complete suite of products (frequency maps, sky models, simulations, beam window functions, time streams, and auxiliary files) is available on the facilities hosted by the National Energy Research Scientific Computing Center (NERSC\footnote{\url{https://www.nersc.gov/}}). We refer to the original paper for a full description and characterisation of what is included in the simulations and how they have been produced. We use the statistics of these simulations to establish an estimation of the \planck{} error-bars $\sigma_{\rm Planck}$ used in Sect.~\ref{sec:comp-Planck} to assess the validity of the models. 

$\sigma_{\rm Planck}$ is derived as the scatter of the power spectra across the 600 NPIPE simulations. We now briefly describe the steps of the pipeline adopted to analyse them:
\begin{enumerate}
    \item Simulated HFI polarised maps at $\nu = [100, 143, 217, 353]$ GHz are loaded in at their original resolution ($n_{\rm side}=2048$);
    \item Maps are degraded to a common resolution of $n_{\rm side}=512$. This step is carried out in spherical harmonics space ($a_{\ell m}$) to avoid introducing artifacts in the output maps;
    \item Cross-split ($A \times B$) and cross-frequency ($\nu_1\times\nu_2$) angular power spectra are estimated with the {\sc Namaster} \citep{namaster} python package, adopting the same configuration as the one described in Sect.~\ref{sec:d12}: bandpowers with constant width of $\Delta\ell = 10$ from $\ell=2$ to $\ell=2n_{\rm side}=1024$, taking into account the mode-coupling due to the adoption of the $f_{sky}=70\%$ galactic mask, correcting for the impact of instrumental beams, and with $B$-modes purification enabled;
    \item For each simulation we end up with $n_{\rm freqs}(n_{\rm freqs}+1)/2 = 10$ power spectra, six cross- and four auto-spectra. Final estimates are the average of these spectra across all 600 simulations, while the scatter $\sigma_{\rm Planck}$ is obtained as the standard deviation. We notice that the latter quantity is not divided by the number of simulations employed for its estimation, as one would do if interested in the error associated with the mean. Instead, we are interested in the spread of the simulations distribution because we want to define distance thresholds considering how much these simulations can scatter.
\end{enumerate}

\section{Power spectra figures \label{app:power-spectra}}
\begin{figure*}
    \centering    \includegraphics[width=\textwidth]{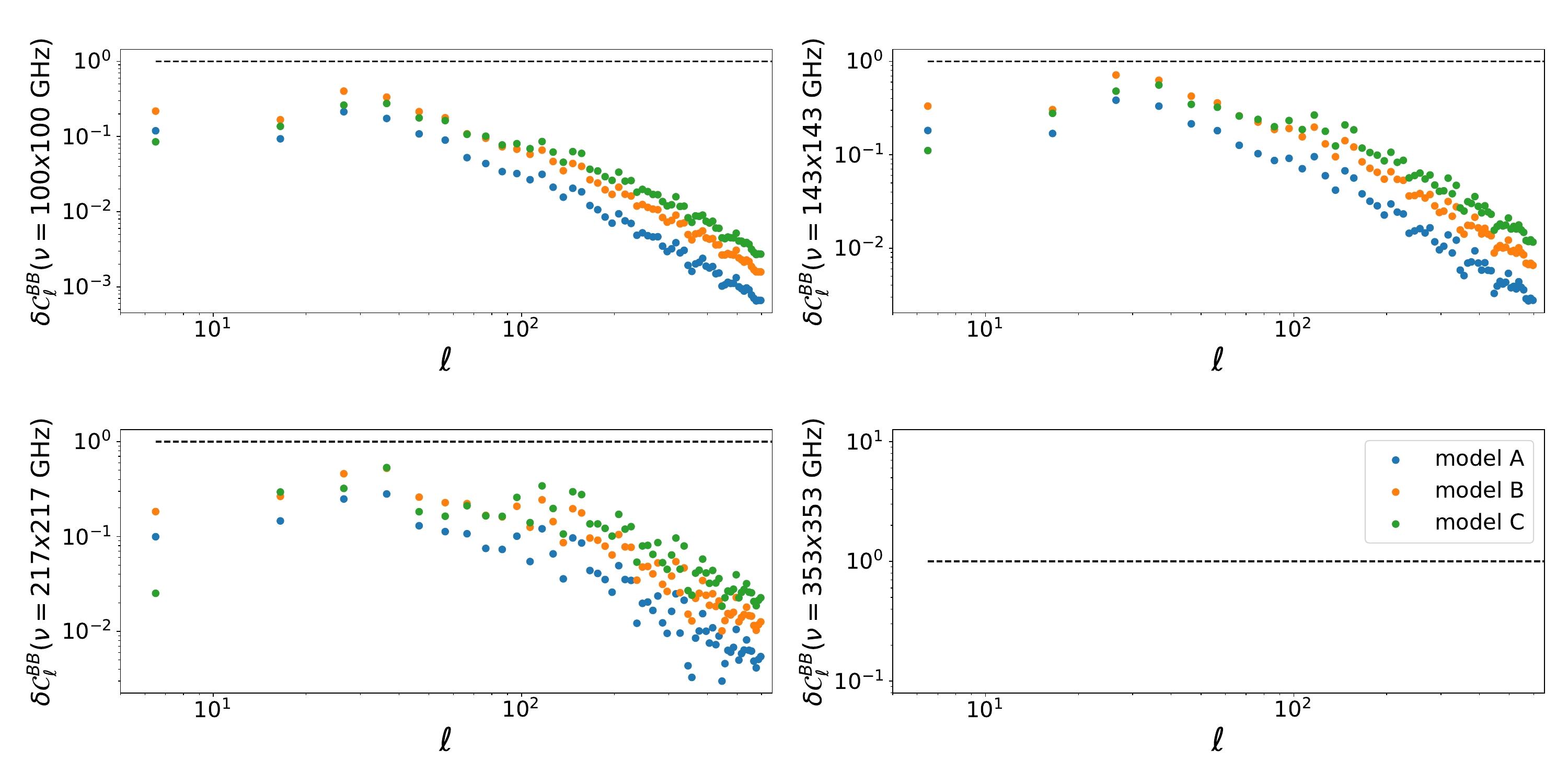}
    \includegraphics[width=\textwidth]{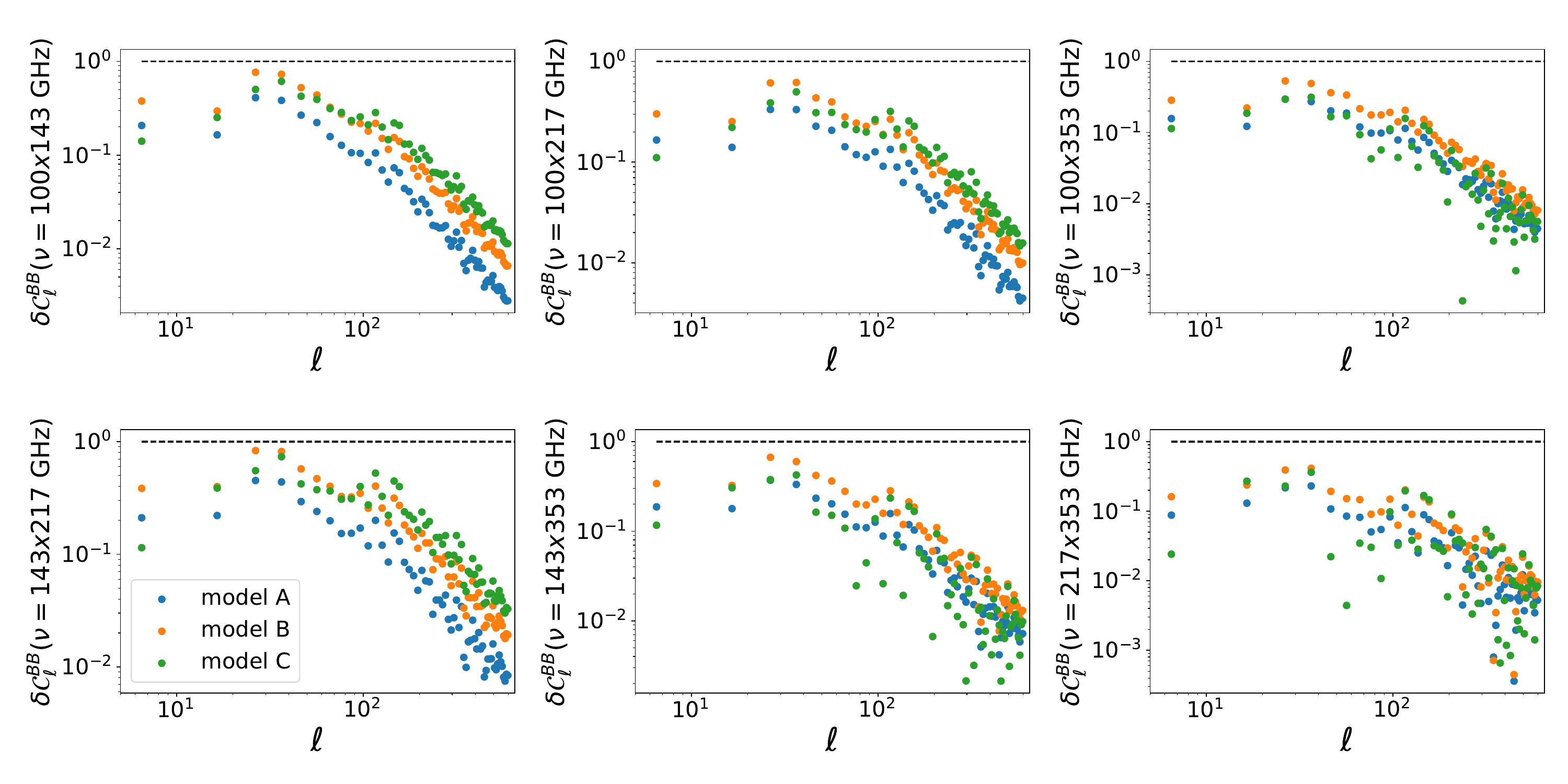}
    \caption{$\delta \mathcal{C}^{BB}_\ell$, as defined in Eq.~\eqref{eq:deltacl}, for the three models and the ten cross-frequency spectra of the \planck{} HFI instrument.}
    \label{fig:model-comp-Planck-BB}
\end{figure*}

\begin{figure*}
    \centering    \includegraphics[width=\textwidth]{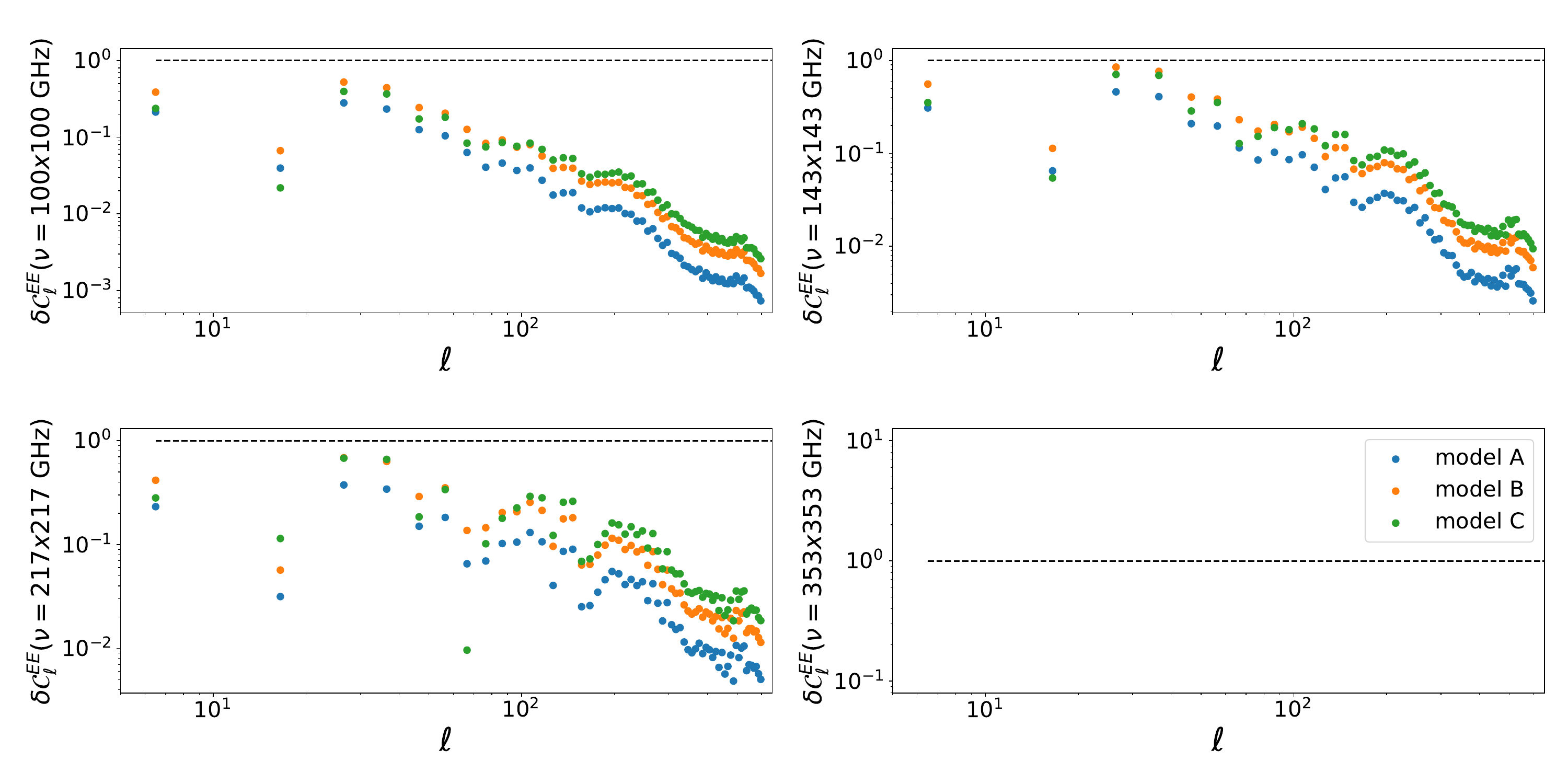}
    \includegraphics[width=\textwidth]{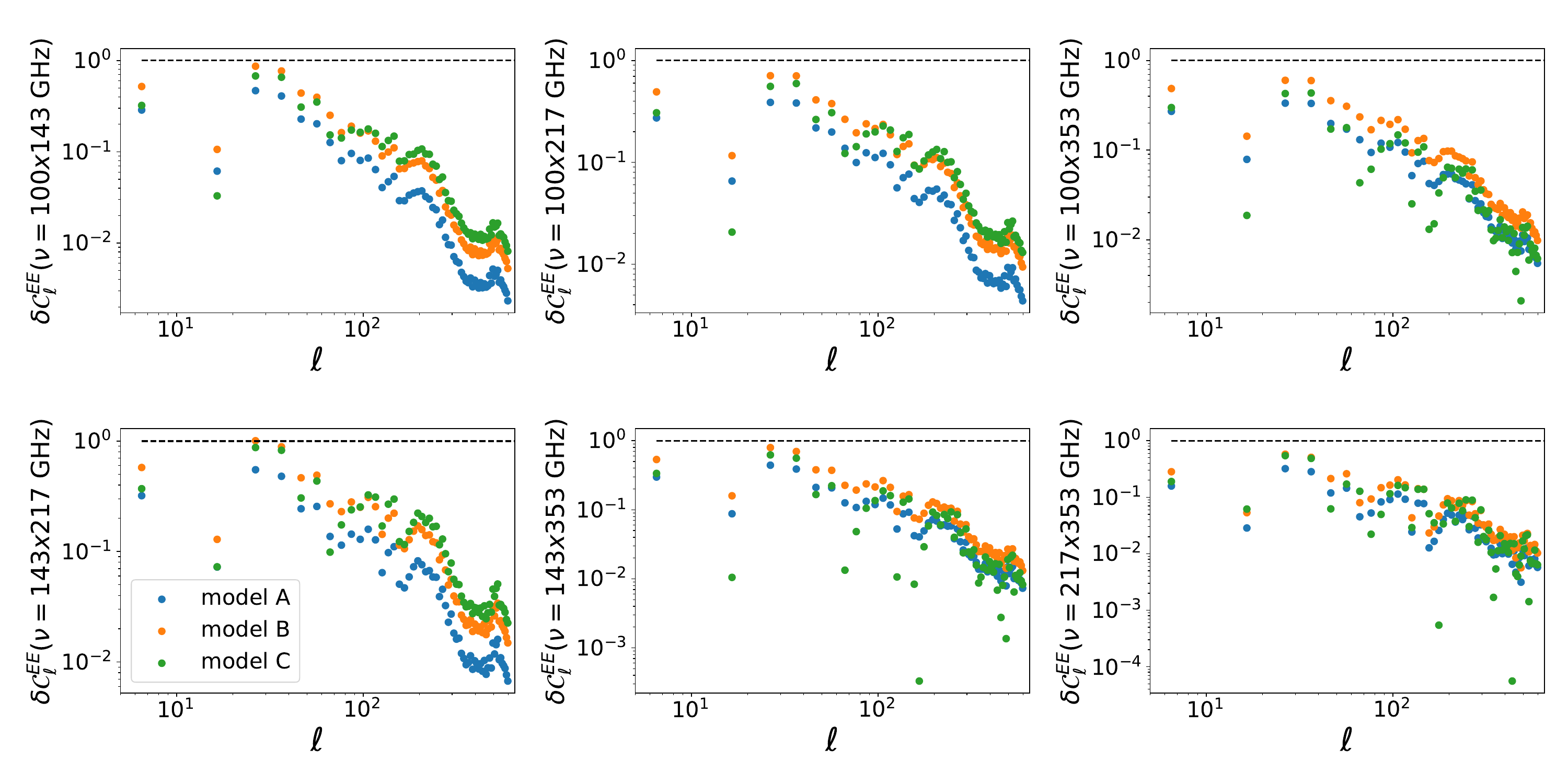}
    \caption{$\delta \mathcal{C}^{EE}_\ell$, as defined in Eq.~\eqref{eq:deltacl}, for the three models and the ten cross-frequency spectra of the \planck{} HFI instrument.}
    \label{fig:model-comp-Planck-EE}
\end{figure*}

\begin{figure*}
    \centering 
    \includegraphics[width=0.97\columnwidth]{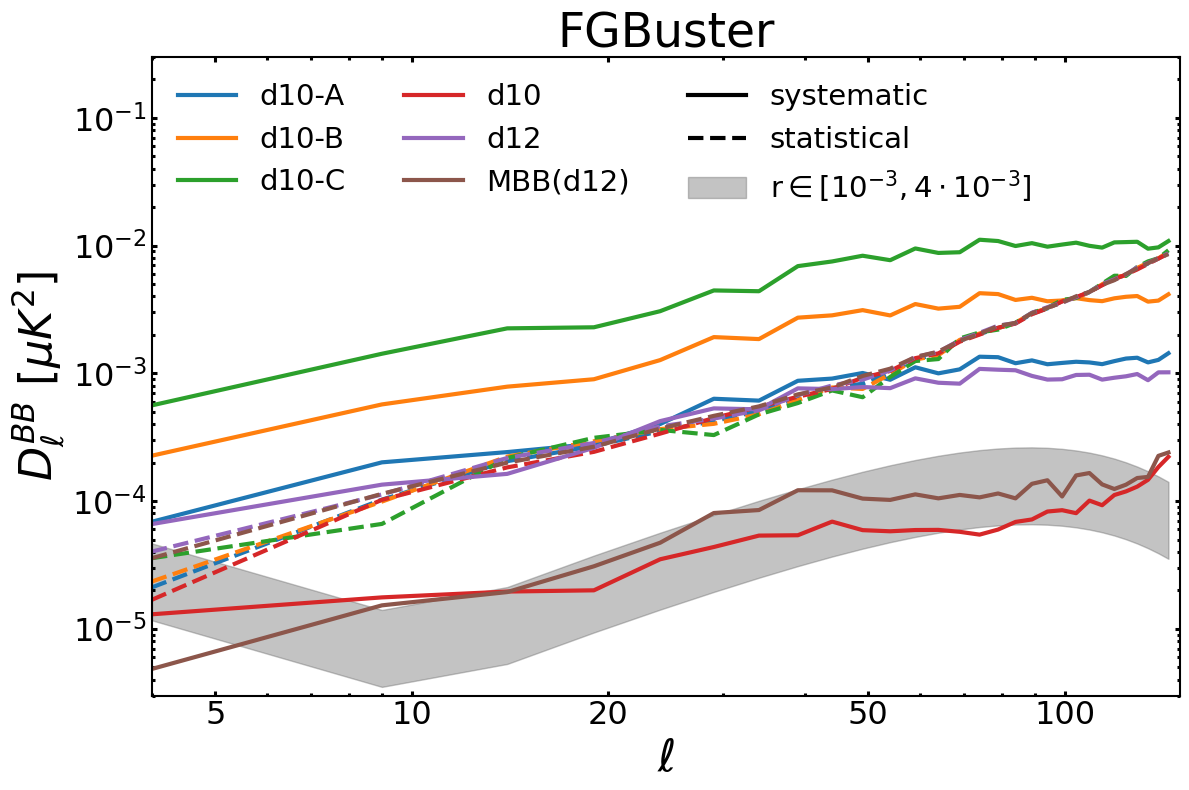}
    \includegraphics[width=0.97\columnwidth]{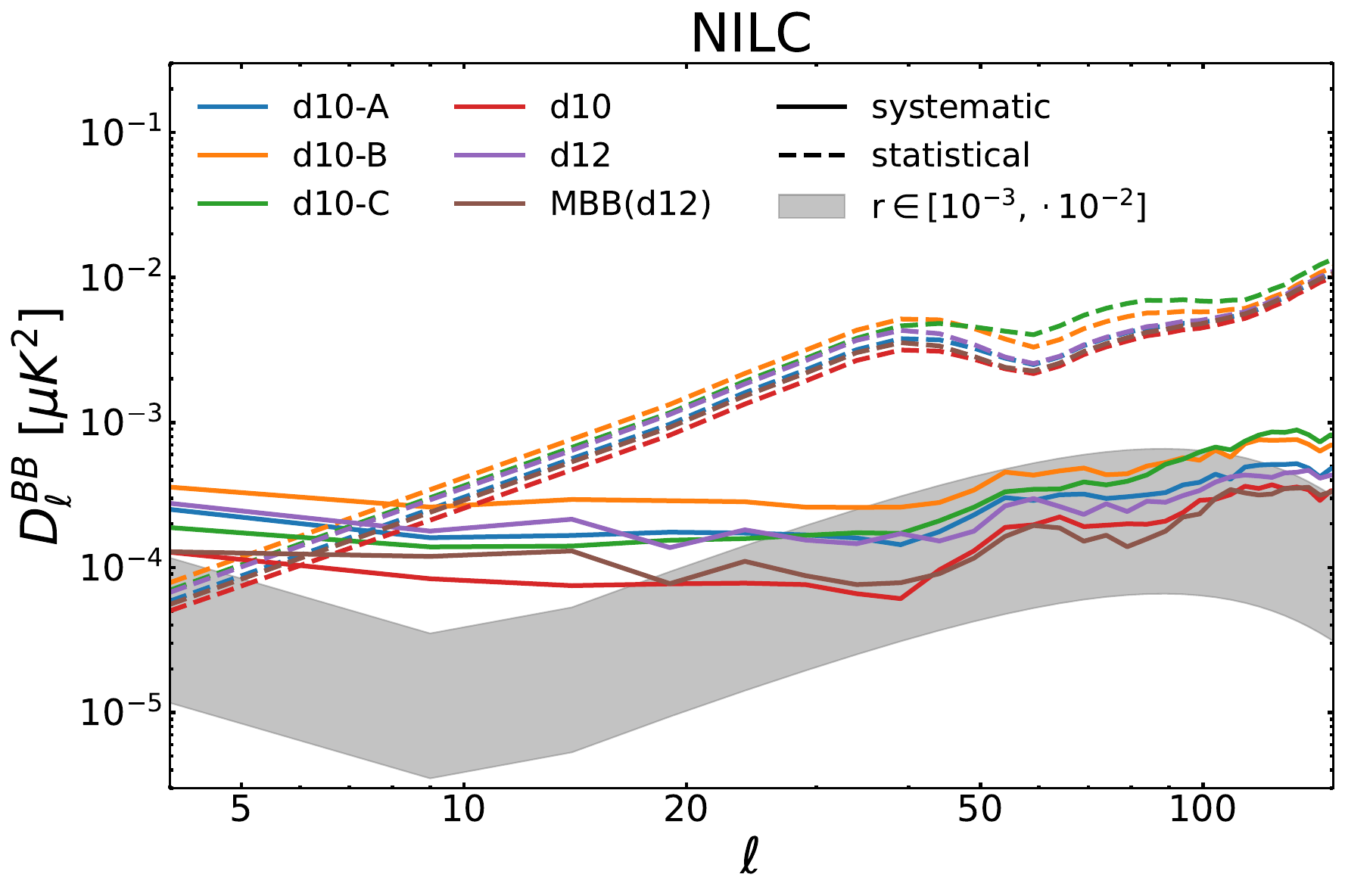}
    \caption{Angular power spectra of $B$-mode residuals in the CMB solutions when using the FGBuster (left) and NILC (right) component separation methods for the different considered foreground models. Systematic residuals are shown with solid lines, while statistical ones with dashed lines. All angular power spectra are computed excluding the Galactic plane with the \planck{} HFI \texttt{GAL70} mask which retains $70\%$ of the sky.}
    \label{fig:comp-sep-full}
\end{figure*}

\begin{figure*}
    \centering    \includegraphics[width=\columnwidth]{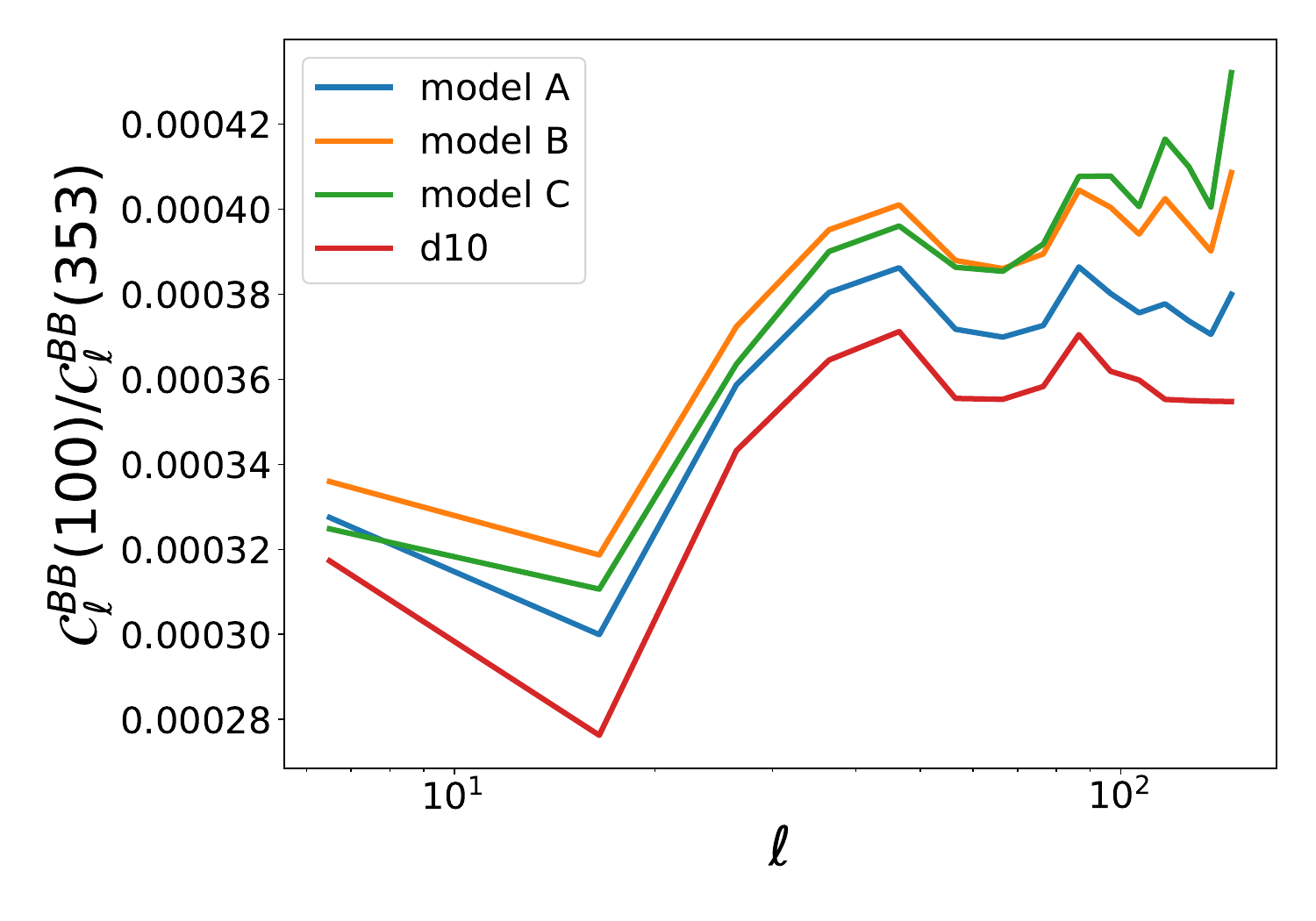} 
    \caption{Ratio of $B$-mode power spectrum of foreground emission at $100$  and $353$ GHz for \dten{} and the three models (A, B, and C) introduced in this work.}
    \label{fig:relative-power}
\end{figure*}
This appendix contains additional figures related to power spectra discussed in the main text. Figures~\ref{fig:model-comp-Planck-BB} and \ref{fig:model-comp-Planck-EE} present $\delta \mathcal{C}_\ell^{XX}$ (as defined in Eq.~\eqref{eq:deltacl}) for the three models considered and the 10 cross-frequency angular power-spectra of the \planck{} HFI instrument.  The full residuals for the two methods (FGBuster and NILC) can be found in Fig.~\ref{fig:comp-sep-full}. Finally, Fig.~\ref{fig:relative-power} present the relative power $BB(100)/BB(353)$ estimated for \dten{} and the three models considered.
\end{appendix}
\end{document}